\documentclass[twocolumn]{aastex631}

\definecolor{cadmiumgreen}{rgb}{0.0, 0.42, 0.24}

\usepackage{comment}
\usepackage{float}
\usepackage{amsmath}
\usepackage{changepage}

\begin{document}

\title{Black Hole Supernovae Outcomes Across a Wide Progenitor Range}

\correspondingauthor{Oliver Eggenberger Andersen}
\email{oliver.e.andersen@astro.su.se}

\author[0000-0002-9660-7952]{Oliver Eggenberger Andersen}
\affil{The Oskar Klein Centre, Department of Astronomy,
Stockholm University, AlbaNova, SE-106 91 
Stockholm, Sweden}

\author[0000-0002-8228-796X]{Evan O'Connor}
\affil{The Oskar Klein Centre, Department of Astronomy,
Stockholm University, AlbaNova, SE-106 91 
Stockholm, Sweden}

\author[0009-0001-5842-7061]{Liubov Kovalenko}
\affil{The Oskar Klein Centre, Department of Astronomy,
Stockholm University, AlbaNova, SE-106 91 
Stockholm, Sweden}

\author[0000-0002-4747-8453]{Haakon Andresen}
\affil{The Oskar Klein Centre, Department of Astronomy,
Stockholm University, AlbaNova, SE-106 91 
Stockholm, Sweden}

\author[0000-0002-5080-5996]{Sean M. Couch}
\affil{Department of Physics and Astronomy, Michigan State University, East Lansing, MI 48824, USA}
\affil{Department of Computational Mathematics, Science, and Engineering, Michigan State University, East Lansing, MI 48824, USA}
\affil{Facility for Rare Isotope Beams, Michigan State University, East Lansing, MI 48824, USA}

\begin{abstract}

Black hole supernovae (BHSNe), the term we use for core-collapse events in which black hole (BH) formation occurs after shock revival but before the explosion is complete, have emerged as a natural outcome of multidimensional simulations as these calculations have been extended to seconds after bounce. Yet they remain one of the least studied outcomes of core collapse. Here, we assess whether they are confined to the most compact and massive progenitors, whose birth rates are low, or whether they arise systematically across a wider range of progenitor structures. We perform 23 long-term axisymmetric core-collapse simulations of progenitors spanning 19.51--60$\,M_\odot$ and compactnesses $0.31 \lesssim \xi_{2.5} \lesssim 0.63$.  We find 18 BHSN outcomes across nearly the full ZAMS mass range considered, corresponding to progenitors with $0.40 \lesssim \xi_{2.5} \lesssim 0.63$. BH formation occurs between $\sim0.7$\,s and $\sim4.4\,s$ after bounce. After BH formation, we continue the evolution with an excision treatment to at least 5000\,s. The final explosion energies span $\sim2\times10^{49}$--$3\times10^{51}$\,erg, while the final BH gravitational masses span $\sim3$--$26\,M_\odot$. We find a clear remnant-mass trend with CO-core mass, but show that the CO core alone is not an adequate proxy for the final BH mass, especially for progenitors at the low- and high-mass ends of the CO-core distribution. Except for the highest CO-core mass models, no single spherical mass coordinate cleanly separates ejecta from remnant material. Finally, a 2D axisymmetric and a 3D model are compared as we discuss differences between the two geometries.

\end{abstract}
 
\section{Introduction} 
\label{sec:intro}
The collapse of the core at the end of a massive star’s life initiates the dynamical sequence that determines the final fate of the star \citep[e.g.][]{bethe:90, janka:12a}. Depending on the progenitor structure, core collapse may produce a neutron star (NS) accompanied by a supernova (SN) explosion, or a black hole (BH) with or without a successful explosion \citep[e.g.][]{heger:03, burrows:25}. Before the outcome is determined, the collapse is abruptly halted by the nuclear force, forming a momentarily stable proto-neutron star (PNS). Consequently, a shock wave launches outward that typically stagnates around $\sim150$\,km \citep[e.g.][]{janka:12b}. If the shock remains stalled, the continued accretion through the shock and onto the PNS from the infalling star causes BH formation, events known as failed SNe \citep{burrows:88, oconnor:11}. The formation of a NS is contingent upon successful shock revival that proceeds to strip the star of most of its mass. This is the conventional SN scenario, but it is not the only possible outcome. A key feature of multidimensional simulations is that a successful shock revival does not necessarily terminate the entire inflow of material onto the central object \citep[e.g.][]{pan:18, kuroda:18, walk:20, powell:21}. Consequently, BHs can form during the early explosion process and impact the explosion dynamics. Recent work has shown that even when a BH forms hundreds of milliseconds after shock revival, the shock can viably survive and blow up the star \citep{chan:18, chan:20, rahman:22, burrows:23, eggenberger_andersen:25, sykes:25, burrows:25}. In \citet{eggenberger_andersen:25}, we referred to these events as black hole SNe (BHSNe), which are perhaps the least studied BH formation channel predicted by core-collapse theory. Regardless of what they are ultimately called, a distinct terminological class is warranted because BHSNe are both dynamically and measurably distinct from fallback SNe in three combined ways: (1) the origin of the accreting material; fallback SNe form black holes from the late-time reversal of ejected matter, whereas in BHSNe, accretion proceeds directly from the infalling star.
(2) the timescale of BH formation; tens of seconds to hours after shock revival in fallback SNe versus hundreds of milliseconds to seconds in BHSNe.
(3) the dynamical consequences; fallback accretion only mildly affects the explosion dynamics, whereas the impact of early BH formation in BHSNe is substantial. These differences also affect the remnant BH mass, the ejecta mass, and the ejecta composition.

If failed SNe, BHSNe, and fallback SNe are all realized in nature as distinct BH formation channels in core-collapse events, it is essential to establish which progenitors lead to each outcome. This is necessary to assess how strongly these channels shape key characteristics of the Universe, including the BH and NS mass distributions, chemical enrichment, and the role of core-collapse SNe in galaxy evolution. These questions are timely, as features in the inferred BH mass distribution are emerging with the growing number of compact-object merger \citep{GWpop4} and X-ray binary \citep{corral-santana:16} observations. In this work, we focus on the viability of BHSN outcomes across a wide progenitor mass range.

Ultimately, the entire progenitor structure needs to be taken into account when investigating the outcome following its core-collapse. However, the compactness as defined in \citet{oconnor:11} is a useful one-parameter diagnostic of the core structure of a star (see Equation~\ref{eq:compactness}). Stars with more compact cores have higher accretion rates after bounce and are therefore more proficient in forming BHs \citep{oconnor:11}. Current state-of-the-art progenitor models and core-collapse simulations do not admit a simple mapping between compactness and successful shock revival. Compactness nevertheless plays a central role in determining the explosion energy, morphology, and heavy-element production \citep{nakamura:15, burrows:24, boccioli:25}. Given that high-compactness progenitors are predisposed to both energetic explosion dynamics and black hole formation, they represent a physically motivated subset in which to investigate BHSNe. Accordingly, previous studies have focused on such models: \citet{chan:18, chan:20} use a 40\,$M_\odot$ zero metallicity progenitor. \citet{burrows:23} use a 40\,$M_\odot$ solar metallically progenitor with a compactness of $\xi_{2.5} = 0.54$. 
\citet{rahman:22} simulate 80\,$M_\odot$ and 115\,$M_\odot$ low-metallicity pulsational pair-instability models with $\xi_{2.5} > 0.77$. \citet{eggenberger_andersen:25} performed three simulations of a solar metallicity 60\,$M_\odot$ progenitor with $\xi_{2.5} = 0.63$. \citet{sykes:25} conducted simulations of eight zero-metallicity progenitors with masses ranging from 60\,$M_\odot$ to 95\,$M_\odot$ and $\xi_{2.5} > 0.62$. In a recent three-dimensional simulation, \citet{burrows:25} found a BHSN outcome for a $19.56$\,$M_\odot$ progenitor ($\xi_{2.5} = 0.45$), which contrasts with the general focus of the BHSN literature on select progenitors above 40\,$M_\odot$, often at low metallicity and higher compactness.

The relation between initial mass and pre-SN compactness depends sensitively on the adopted stellar evolution model. The outcome can be strongly influenced by binarity \citep[e.g.][]{schneider:21}, stellar mass-loss rates \citep[e.g.][]{sukhbold:18}, rotation \citep[e.g.][]{limongi:18}, nuclear reaction rates \citep[e.g.][]{tur:07}, and three-dimensional effects, such as those impacting shell interactions \citep{whitehead:26}. Despite remaining uncertainties in stellar evolution, it is essential to investigate how a range of progenitor structures affects the post-core-collapse evolution and final outcome. While BHSNe have been demonstrated as a possible outcome in two and three-dimensional simulations, their possible impact in the Universe remains unclear since most of the models in the literature are of extremely high-mass progenitors with rare birth rates. In this study, we explore their potential impact by performing long-term simulations of 23 single stars with zero-age main sequence (ZAMS) masses between $\sim19-60$\,$M_\odot$, with $ 0.31 \lesssim \xi_{2.5} \lesssim  0.63$. As a consequence of evolving the simulations through the entire accretion process, we find 18 BHSN outcomes across the entire ZAMS mass range, corresponding to compactnesses between $ 0.40 \lesssim \xi_{2.5} \lesssim  0.63$. The final explosion energies vary between $\sim 2\times10^{49}$ and $\sim 3\times 10^{51}$\,erg. We find gravitational masses of the BH remnants between $\sim 3\,M_\odot$ and $\sim 26$\,$M_\odot$, which generally increase with ZAMS mass, primarily reflecting the growth of the CO core with progenitor mass in this model set. We further demonstrate that no single progenitor mass coordinate cleanly partitions the star into ejecta and remnant mass, except in models with the most massive CO cores.

\section{Methods}
\label{sec:methods}

The simulations are performed using FLASH (v4) \citep{fryxell:00, dubey:09}, modified for core-collapse SNe \citep{couch:13a, couch:14a, oconnor:18, oconnor:18c}. The simulation procedure is described in two stages: (1) the evolution up to BH formation, which occurs between $\sim0.65$\,s and $\sim4.35$\,s after bounce depending on the model; and (2) the evolution after BH formation, where we apply an excision mask around the BH to follow both the explosion and the continued accretion onto the BH. We evolve the explosions to at least a 5000\,s, by which time the explosion and remnant properties have largely converged to their asymptotic values.

\begin{figure}[ht!]
    \centering
    \includegraphics[width=\columnwidth]{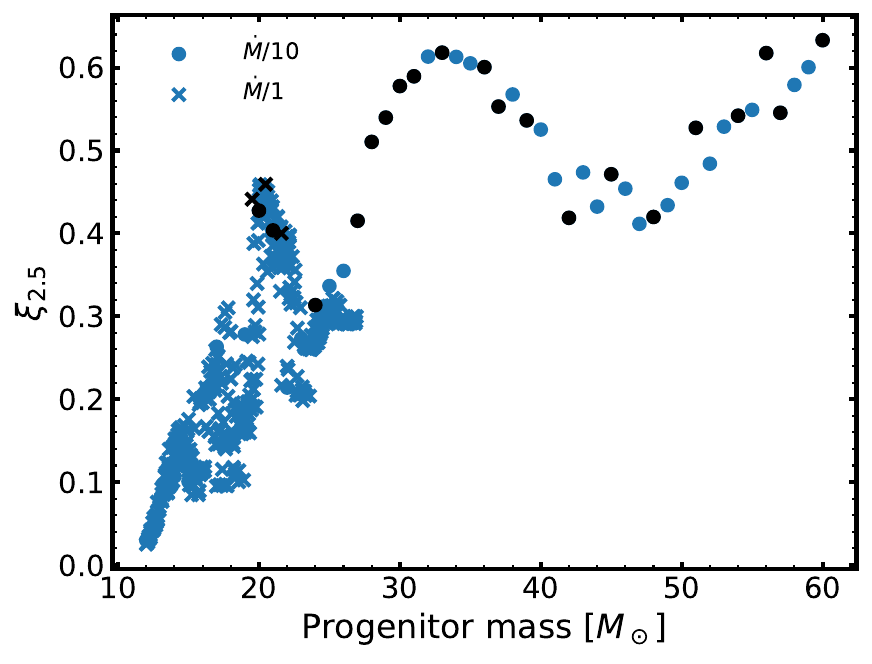}
    \caption{Compactness $\xi_{2.5}$ as a function of ZAMS mass for the progenitor suite of \citet{sukhbold:18}. Crosses indicate models evolved with the nominal mass-loss rate, while dots indicate models evolved with one-tenth of this rate. Black markers highlight the 23 progenitors selected for this study, spanning $19.51$–$60$\,$M_\odot$ and a compactness range of $0.31 \lesssim \xi_{2.5} \lesssim 0.63$. The sample covers the first compactness peak near $20$\,$M_\odot$, where the two sets overlap, as well as the second and third peaks in the reduced mass-loss set.} 
    \label{fig:compactenss_sample}
\end{figure}

\begin{figure*}[ht!]
    \centering
    \includegraphics[height=\columnwidth, width=1.5\columnwidth]{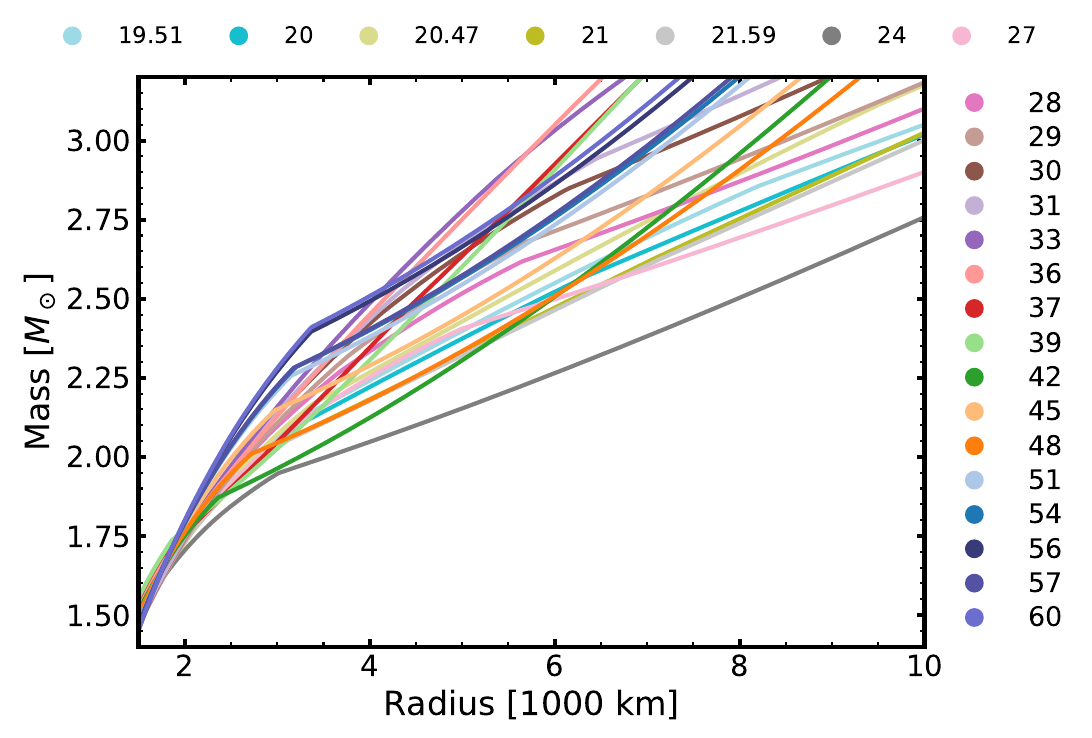}
    \caption{Mass--radius relations for the progenitor suite, showing enclosed mass out to 3.25\,M$\odot$ as a function of radius, color-coded by the progenitor ZAMS mass. The models are nearly equally compact below 1.75\,M$\odot$, but diverge beyond this point. While shock revival impacts the accretion rate, the underlying mass--radius relation remains closely linked to the subsequent accretion rate, PNS mass growth, BH formation time, and explosion energy.} 
    \label{fig:mass-radius}
\end{figure*}

\subsection{Evolution before black hole formation}
The simulations are evolved in two dimensions, assuming axisymmetry in a cylindrical coordinate system. The computational domain extends to $\pm1\times10^{10}$\,cm along the axial direction and $1\times10^{10}$\,cm in radius. It is initially divided into $16\times8$ grid blocks, each containing $16\times16$ zones. Depending on the model and evolutionary stage, 12, 13, or 14 levels of adaptive mesh refinement are employed, corresponding to a finest spatial resolutions of approximately 381\,m, 191\,m, and 95\,m, respectively. For PNSs older than $\sim1.5$\,s, steep density gradients develop at their outer edges, which can become insufficiently resolved at 381\,m and, at later times, even at 191\,m resolution. We discuss our investigation on the impact of resolution on the numerical solution in Section~\ref{sect:resolution}. The grid is maximally refined up to 90\,km, beyond which refinement criteria based on gradients in density and pressure are applied, with a ceiling corresponding to an effective angular resolution of $\,0.6^\circ$. By this, we mean that blocks with $\Delta x/r < 0.6^\circ \times \pi/180^\circ$ are prevented from refining further.

Hydrodynamics is evolved in the Newtonian limit, with general-relativistic effects incorporated through an effective relativistic gravitational potential (the ``Case~A" potential of \citealt{marek:06}), which has been shown to reproduce the key relativistic effects relevant for core-collapse dynamics \citep{marek:06, mueller:12a, oconnor:18}. Comparisons between one-dimensional simulations with this potential and fully relativistic treatments show good agreement in black-hole formation times \citep{schneider:20}. Time integration is performed with a second-order Runge–Kutta scheme \citep{couch:21}. Spatial reconstruction uses the WENO-Z algorithm \citep{borges:08}, and fluxes across cell interfaces are computed using the HLLC Riemann solver \citep{toro:94}. The HLLE solver \citep{HLLE:88} is adopted in regions containing shocks. The equations of hydrodynamics are closed using a hybrid EOS \citep{witt:21}. At high densities, the SFHo nuclear equation of state is used \citep{steiner:13a}, and at low densities it is transitioned to the $Y_e$ based Helmholtz EOS \citep{timmes:00}.

For neutrino transport, we evolve the first two angular moments of the neutrino distribution function, with the system closed by the analytic M1 approximation to the full angular dependence \citep{shibata:11,oconnor:18}. The energy domain is discretized into 12 logarithmically spaced bins from 1 MeV to about 315 MeV, for three neutrino species: electron neutrinos ($\nu_e$), electron antineutrinos ($\bar{\nu}_e$), and the combined heavy-lepton neutrino group ($\nu_x$). Neutrino interaction rates are computed with \texttt{NuLib} \citep{oconnor:15a} using the baseline opacity set of \citet{oconnor:18c}, supplemented with mean-field and virial corrections \citep{horowitz:2017}, final state blocking for both absorption and scattering processes, as well as inelastic scattering on electrons following \citet{bruenn:85}.

\subsection{Evolution after black hole formation}

We largely follow the methods outlined in \citet{eggenberger_andersen:25}. At BH formation, we excise the central region \citep[similar to, e.g.][]{chan:18, rahman:22, schneider:23, sykes:23, vartanyan:25, burrows:25} and continue the evolution for at least 5000\,s, by which time the global explosion and remnant properties have largely converged to their asymptotic values. The excision mask is initialized at 100\,km and expands at 100\,km\,s$^{-1}$. Inside the mask, the density, temperature, and internal energy are set to a small fraction of their values just outside it, thereby maintaining a low-pressure region that approximates an absorbing inner outflow boundary. Mass entering the excised region is added to a central point mass, whose estimated gravitational mass is used in a Newtonian gravitational potential. The gravitational mass ($M_\mathrm{grav}$) of the compact object at time $t$ is calculated by subtracting the mass-energy radiated away by neutrinos from the baryonic mass ($M_\mathrm{bary}$): 
\begin{equation}\label{eq:mgrav}
M_\mathrm{grav}(t) = M_\mathrm{bary}(t) - \frac{1}{c^2}\sum_\nu \int_0^t L_\nu(t^\prime)dt^\prime,
\end{equation}
where $L_\nu$ is the neutrino luminosity from species $\nu$. The point mass at the time of BH formation ($t_\mathrm{BH}$) is obtained by inserting $t=t_\mathrm{BH}$ into Equation~\ref{eq:mgrav}. The neutrino luminosity is expected to drop by 1–2 orders of magnitude on a millisecond timescale after BH formation \citep{rahman:22, sykes:23, sykes:25}, rendering the impact of neutrinos after BH formation insignificant, and we therefore discontinue the transport at BH formation.

At BH formation, we enlarge the computational domain by constructing a larger grid filled with the progenitor data and mapping the simulated data onto its inner region. The enlarged domain extends to $\pm1.024\times10^{13}$\,cm along the axial direction and $1.024\times10^{13}$\,cm in radius. The domain is also populated by 180000 passive particles logarithmically spaced up to $1\times10^{11}$\,cm in radius and evenly spaced in $\cos\theta$, where $\theta$ is the polar angle. The particles follow the advection streams and allow us to pinpoint which material in the domain at BH formation eventually falls into the BH and which escapes.

\begin{figure*}[ht!]
    \centering
    \includegraphics[height=\columnwidth, width=1.5\columnwidth]{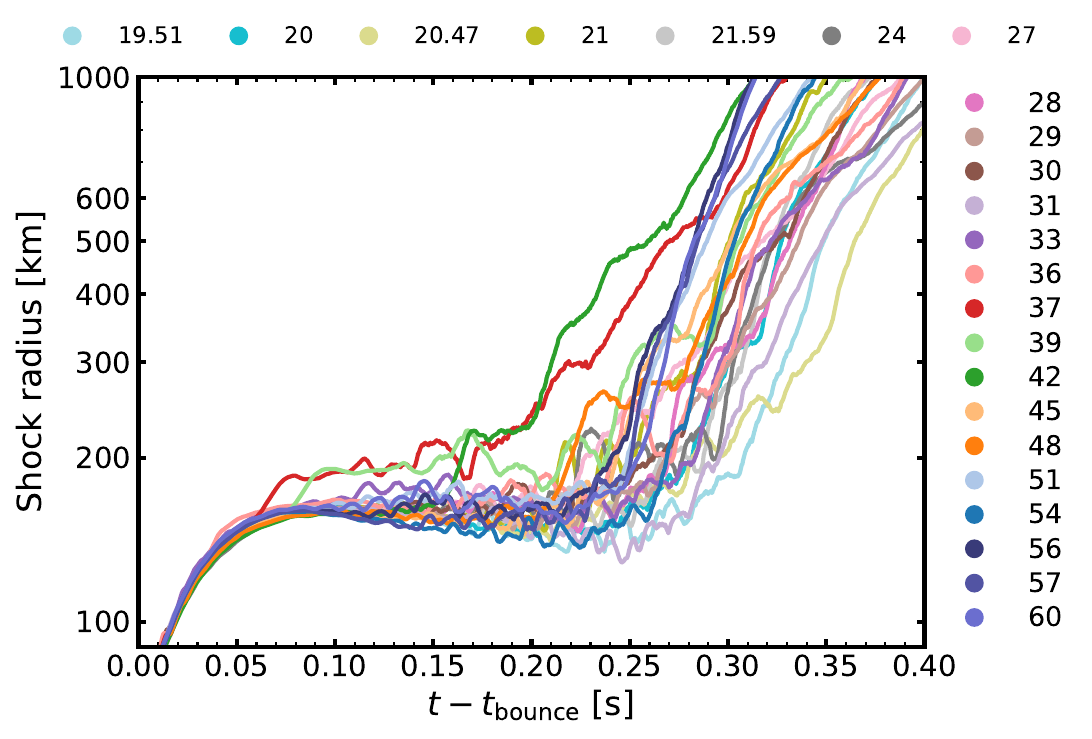}
    \caption{Mean shock radius evolution. All models undergo shock revival, which we attribute to the high neutrino-heating rate in high-compactness models, the SFHo EOS used, and possibly the 2D axisymmetric constraint. Shock expansion triggered by the accretion of density jumps is evident in some models, and occurs especially early in the 37\,$M_\odot$ and 39\,$M_\odot$ models, both of which later produce remarkably energetic BHSNe.} 
    \label{fig:shock_radius}
\end{figure*}

\subsection{Progenitors}\label{sect:progenitor}

We use 23 progenitors from \citet{sukhbold:18} in the ZAMS mass range of 19.51--60\,$M_\odot$, corresponding to a compactness range of $0.31  \lesssim \xi_{2.5} \lesssim  0.63$, where the compectness is defined as \citep{oconnor:11}:

\begin{equation}\label{eq:compactness}
\xi_{M} = \frac{M/M_\odot}{R(M_{\mathrm{bary}} = M)/1000\,\mathrm{km}}
\end{equation}.

Figure~\ref{fig:compactenss_sample} shows the progenitors plotted by compactness and ZAMS mass, where black markers indicate the stars used in this study. Crosses denote stars evolved with a nominal mass-loss rate (models 21.59\,$M_\odot$, 20.47\,$M_\odot$, and 19.51\,$M_\odot$), while dots denote stars evolved with a tenth of the nominal rate. The sample includes stars around the first compactness peak at $\sim20$\,$M_\odot$, a region where the effect of reduced mass-loss rate on the core structure is small, and the two sets overlap \citep{sukhbold:18}. The reduced mass-loss suite allows us to probe progenitors around a second and a third compactness peak. 

The 2.5\,$M_\odot$ mass coordinate is a useful choice in the compactness parameter for studies of BH formation, because the PNS baryonic mass at collapse is often near this value. In failed SNe, where the shock is not revived, the BH formation time is tightly correlated with \(\xi_{2.5}\) \citep{oconnor:11, schneider:20}. In BHSNe models, the relation is more complex due to shock revival. The shock expansion reduces the accretion rate and prolongs the time to BH formation. Moreover, because explosion and accretion occur simultaneously, mass shells beyond 2.5\,$M_\odot$ can accrete onto the PNS before BH formation and thus remain relevant as fuel for the explosion. Figure~\ref{fig:mass-radius} shows the enclosed mass out to 3.25\,$M_\odot$ as a function of radius. Although shock revival and multidimensional effects introduce additional nuances, this mass--radius relation still correlates with the accretion rate, and therefore with the PNS mass evolution, BH formation, and the explosion energy, as will be evident in Section~\ref{sec:results}. The intersections of the curves with a horizontal line at a given mass coordinate define the spread in compactness at that coordinate. These progenitors are nearly equally compact below 1.75\,$M_\odot$, but diverge beyond that point.

\section{Results}\label{sec:results}

\subsection{Shock revival}
Figure~\ref{fig:shock_radius} shows the mean shock-radius evolution of all the models during the first 0.4 seconds post bounce. All models undergo shock revival, which we attribute to the high neutrino-heating rate in high-compactness models \citep{burrows:24, boccioli:25}, the relatively soft SFHo EOS used \citep[e.g.][]{schneider:20, boccioli:24review}, and possibly the 2D axisymmetrical constraint. Across the suite, the mean shock radii reach 400\,km between $\sim0.23$\,s and $\sim0.36$\,s after bounce. We highlight the relatively early shock expansion for models 37\,$M_\odot$ (light green) and 39\,$M_\odot$ (dark red), which we later argue can be a contributing factor to their remarkably energetic outcomes (see Section~\ref{sect:explosion_dynamics}). Their early shock expansion occurs due to the accretion of a sharp compositional interface and a corresponding drop in the density profile. 

Although the shock expansion dampens the accretion rate, it does not necessarily reverse the direction of all the overlying material \citep{mueller:15b, powell:21}, a feature that is especially prominent for compact progenitors evolved in multiple dimensions \citep{burrows:24}. Such is the case for the simulations we present here; generally, the explosions develop preferentially along some directions such that the accretion-flow funnels into the other directions. The preferential explosion direction can be seeded by the standing accretion shock instability \citep{blondin:03}, with an effective bias of the large-scale outflows towards the poles in axisymmetric simulations \citep{janka:12b, mueller:15b}. 

\begin{figure*}[ht!]
    \centering
    \includegraphics[height=\columnwidth, width=1.5\columnwidth]{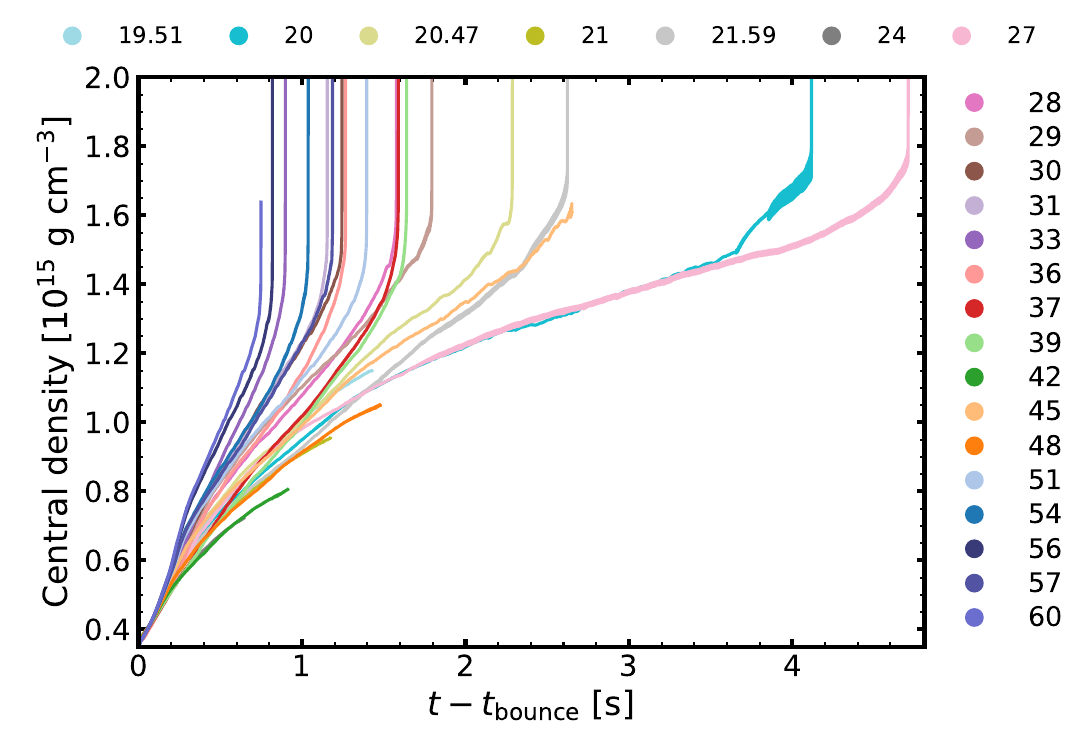}
    \caption{Central density evolution of the PNSs. Collapse to a BH is marked by a sudden rapid rise in central density. 18 out of 23 models form BHs between $\sim0.7$--$4.1$\,s after bounce. The central density at BH formation lies within the range of $1.3\times10^{15}$\,g\,cm$^{-3}$ to $1.8\times10^{15}$\,g\,cm, with the upper end reached by models that form BHs late. The lower-density models are discontinued before 2\,s, as they appear less likely to form BHs through continued infall from core collapse, although several additional seconds of evolution would be required to determine this conclusively.} 
    \label{fig:central_density}
\end{figure*}
\subsection{PNS evolution and BH formation}

Figure~\ref{fig:central_density} shows the central density evolution of the PNSs, which continues to increase as the PNSs contracts by mass growth, deleptonization, and cooling. The onset of collapse to a BH is marked by a rapid increase in central density on a millisecond timescale. The BH formation times span a wide range, from \(\sim0.7\,s\) to \(\sim4.4\,\mathrm{s}\) after bounce. The central density at the onset of rapid collapse also tends to be higher for later-forming BHs, increasing from $1.3\times10^{15}$\,g\,cm$^{-3}$ to $1.8\times10^{15}$\,g\,cm$^{-3}$. We note that this trend agrees with the Tolman–Oppenheimer–Volkoff solutions for the maximum mass PNS configurations using the SFHo EOS, given that older PNSs have evolved to a lower average entropy (more details further down in this section). The sharp bend in central density for the 20\,$M_\odot$ simulation around 3.6\,s is numerical in nature and arises when the PNS relaxes into a more compact configuration after we increase the resolution. Because we expect the true BH formation time in this model to be close to the time of the simulated PNS collapse, we retain it in the figures. We avoid such numerical artifacts in the other models by increasing the resolution at an earlier stage. We discuss these challenges of late-time PNS modeling in Section~\ref{sect:resolution}.

\begin{figure*}[ht!]
  \centering
  \begin{adjustwidth}{-0.8cm}{0.4cm}
    \includegraphics[width=\dimexpr\textwidth+1.2cm\relax]{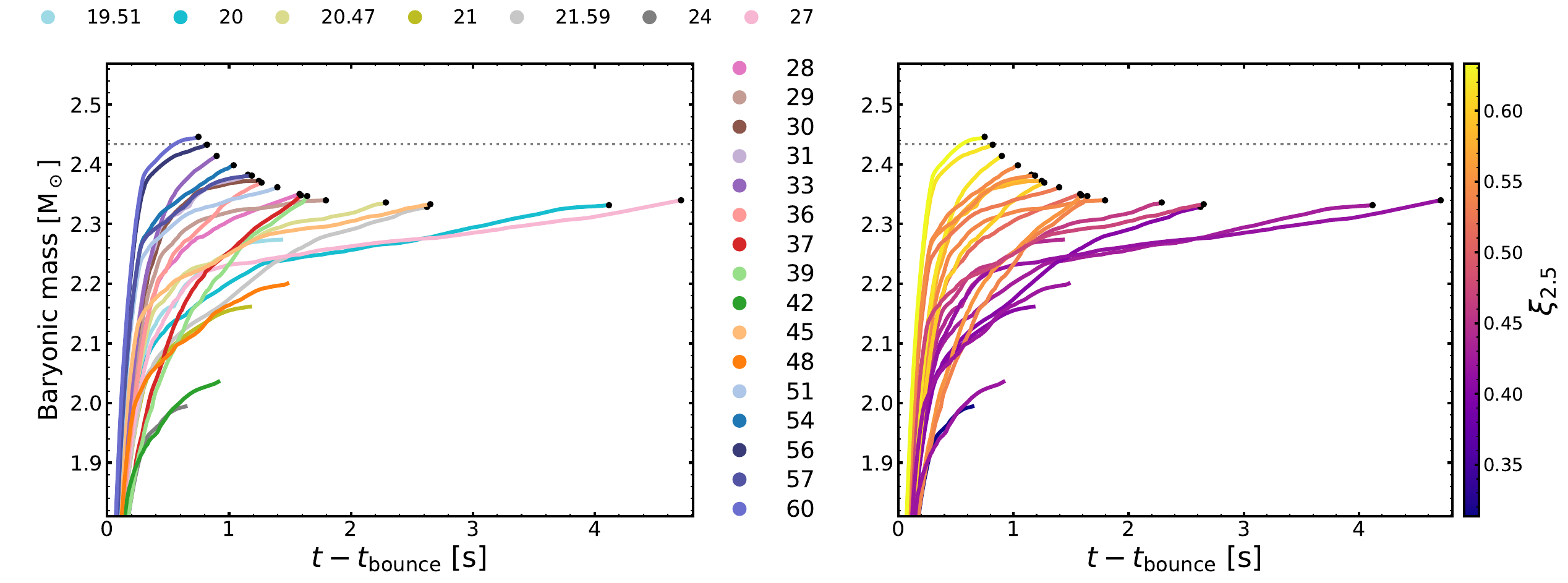}
  \end{adjustwidth}
  \caption{The baryonic mass evolution of the PNSs, explicitly labeled by each progenitor ZAMS mass (left) and by compactness (right). Black dots indicate the moment of BH formation, which generally correlates with the compactness. However, predicting the detailed path to BH formation and the growth rate of the PNS, requires taking into account the entire inner mass--radius relation of the progenitor (see Figure~\ref{fig:mass-radius}). Most PNSs collapse to a BH below the maximum mass of a cold NS (horizontal dotted line) since their thermal energy content contributes more to gravity than to supporting the star (see text and Figure~\ref{fig:mass_entropy}).}
  \label{fig:M_NS}
\end{figure*}

Figure~\ref{fig:M_NS} shows the baryonic mass evolution of the PNSs. In the left panel, all simulations are explicitly labeled; in the right panel, simulations are color-coded by their compactness at 2.5\,$M_\odot$ ($\xi_{2.5}$). The black dot at the end of lines indicates the moment of BH formation, and the horizontal line is the maximum baryonic mass of a cold NS for the SFHo EOS. Models that do not appear to reach BH formation through infall accretion are not evolved beyond 2\,s. However, confirming this would require evolving these models for several additional seconds. The 19.51\,$M_\odot$ model is not such a case, as it appears to be evolving towards BH formation, but this model is discontinued due to a crash. As mentioned in Section~\ref{sect:progenitor}, the details in the mass growth rate of the PNS is attributed to the mass--radius structure of the progenitor (Figure~\ref{fig:mass-radius}). Mapping the mass–radius relation at bounce to a free-fall time for each mass shell yields a curve that closely tracks the PNS baryonic-mass evolution up to shock revival. Shock revival partially ejects material, reducing the accretion rate and slows the inflow. Moreover, the neutrino engine is quite efficient \citep{bollig:21}, even after $\sim$1--2\,s, we observe that a significant fraction ($\sim$10--50\,\%) of infall material can be sent back up in neutrino-driven winds. These two combined effects flatten the slopes in Figure~\ref{fig:M_NS} and significantly prolongs the time to BH formation compared to a failed SN without shock revival. Despite this, there remains a strong correlation between the compactness and the BH formation time. The 60\,$M_\odot$ progenitor, which is the most compact star in this sample, forms a BH first. More generally, the BH formation time tends to increase with decreasing compactness. Since the compactness is not a monotonic function of the ZAMS mass (see Figure~\ref{fig:compactenss_sample}), neither is the BH formation time. The third fastest BH formation comes from a 33\,$M_\odot$ star, which the ``second compactness" peak is centered around (see Figure~\ref{fig:compactenss_sample}). Between the second and the third compactness peaks are some relatively high-mass progenitors that, based on our simulations, will likely not form BHs from the infall material (but perhaps from later fallback). Such examples are the 42\,$M_\odot$ star (green line) and the 48\,$M_\odot$ star (orange line). Despite the correlation between BH formation time and compactness, the PNS mass-evolution curves are not cleanly ordered and often intersect, because the progenitor mass–radius relations are themselves not cleanly ordered, and because of stochastic effects. In a later section, we exemplify how the path to BH formation matters for the explosion development and dynamics. Finally, we note the curved locus traced by the black dots in Figure~\ref{fig:M_NS}, which shows the relation between the PNS baryonic mass and the BH-formation time. The existence of this locus is nontrivial and reflects the dependence of the PNS maximum mass on its interior thermodynamic structure \citep{schneider:20}, which we now highlight in more detail.

\begin{figure}[ht!]
    \centering
    \includegraphics[width=\columnwidth]{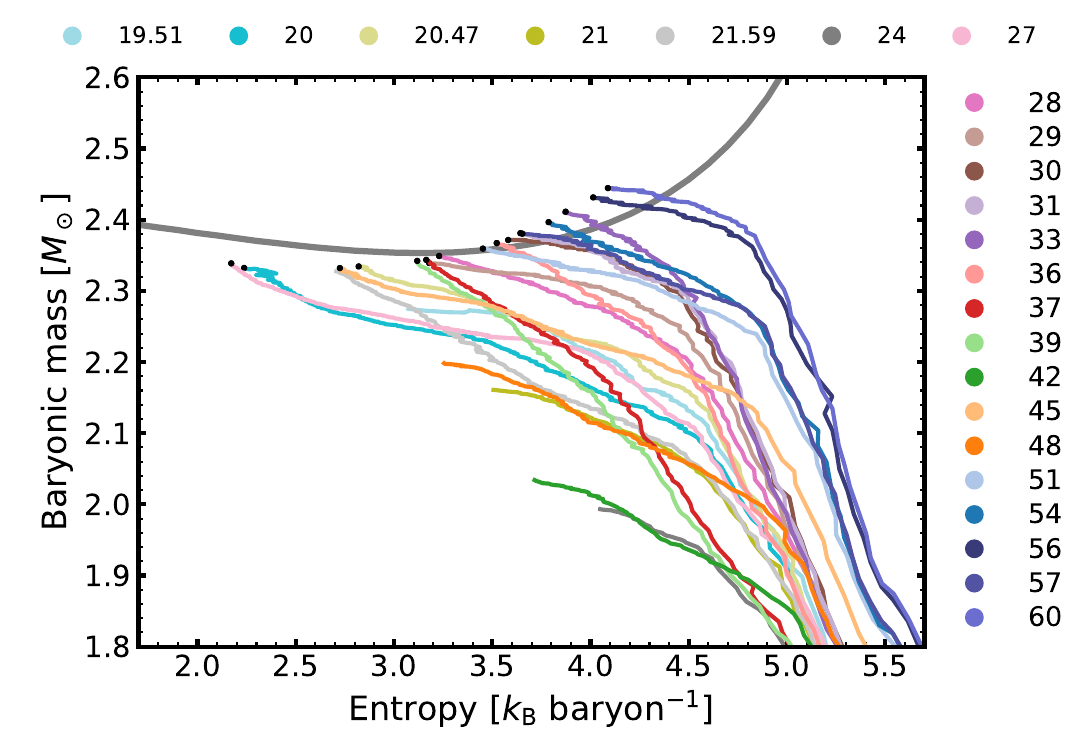}
    \caption{Gravitational mass of the PNS as a function of its entropy (see text for definition, colored lines), with black dots indicating BH formation. The gray U-shaped line represents the maximum baryonic mass, $M^{\mathrm{max}}(s)$, of a spherically symmetric, constant entropy NS, assuming the SFHo EOS. Despite PNS entropy profiles that are not constant in the simulations, BH formation occurs close to the general relativistic prediction \citep{schneider:20}. Therefore, there is a correspondence between the functional form of the maximum mass and the curved line that traces the locus of points in Figure~\ref{fig:M_NS}.} 
    \label{fig:mass_entropy}
\end{figure}

The entropy ($s$) in the PNS is a rough proxy for the thermal energy stored in the PNS. Following the work of \citet{schneider:20}, the gray line in Figure~\ref{fig:mass_entropy} shows the PNS maximum baryonic mass versus entropy, $M^{\mathrm{max}}(s)$, calculated via the Tolman–Oppenheimer–Volkoff equations assuming a constant-entropy NS and the SFHo EOS. \(M^{\mathrm{max}}(s)\approx 2.43\,M_\odot\) at zero entropy, decreases to a minimum of \(\sim2.35\,M_\odot\) near \(s\approx3.1\,k_\mathrm{B}\,\mathrm{baryon}^{-1}\), and then rises steeply for \(s\gtrsim4\,k_\mathrm{B}\,\mathrm{baryon}^{-1}\). At high entropies, the thermal energy contributes more to the pressure support than to gravity, increasing the maximum mass. At intermediate entropies near the minimum, the opposite is true: thermal energy contributes more to gravity than to pressure support, reducing the maximum mass relative to both a cold NS and a very hot PNS. The existence of this minimum implies that any PNS with a mass above \(\sim2.35\,M_\odot\) and an entropy above \(s\sim3.1\,k_\mathrm{B}\,\mathrm{baryon}^{-1}\) will inevitably collapse to a BH as it cools, even without further accretion. In the context of failed SNe, \citet{schneider:20} mapped the evolution of PNSs onto the mass--entropy diagram by tracking both their mass and a representative entropy measure. They showed that BH formation occurs when the evolutionary tracks, which are straight lines in 1D, come into close proximity to \(M^{\mathrm{max}}(s)\).

For the simulations presented here, the colored lines in Figure~\ref{fig:mass_entropy} show the evolution of the PNSs in the baryonic mass--entropy diagram. We define a representative entropy by first constructing a histogram of mass as a function of entropy, then identifying the entropy bin containing the largest mass, and finally computing the mass-weighted average entropy within a window of width 1.2\,$k_\mathrm{B}$\,baryon$^{-1}$ centered on that bin. Similar to the failed-SN case \citep{schneider:20}, higher-compactness models trace higher-entropy trajectories in the mass--entropy diagram. For example, the most compact progenitor is the 60\,$M_\odot$ star, which is the rightmost (purple) line in the diagram, while the 42\,$M_\odot$ progenitor located between the second and the third compactness peaks is far to the left (green line). Lines from different progenitors cross however, and the detailed trajectory is again sensitive to the mass--radius relation of the progenitor (Figure~\ref{fig:mass-radius}), plus stochastic effects. A notable difference compared to failed SNe, however, is the bent mass--entropy curves, which is a consequence of the altered accretion rate following shock revival. Nevertheless, the BH formation occurs rather close to $M^{\mathrm{max}}(s)$. The curved locus traced by the black dots in Figure~\ref{fig:M_NS} is a reflection of the curved locus in Figure~\ref{fig:mass_entropy}. The relatively small deviation between the locus and $M^{\mathrm{max}}(s)$ can arise because the simulated PNSs are not constant in entropy, and the neutrino fields are not in neutrino-free beta equilibrium (as assumed for the constant entropy TOV stars). Furthermore, deviations from full GR can arise due to the use of an effective potential, which may be particularly important at late times in the lowest compactness BHSNe. We remark that we have run even higher compactness models, and see a closer agreement between the locus and $M^{\mathrm{max}}(s)$ again at higher entropies.

\begin{figure*}[ht!]
  \centering
  \begin{adjustwidth}{-0.8cm}{0.4cm}
    \includegraphics[width=\dimexpr\textwidth+1.2cm\relax]{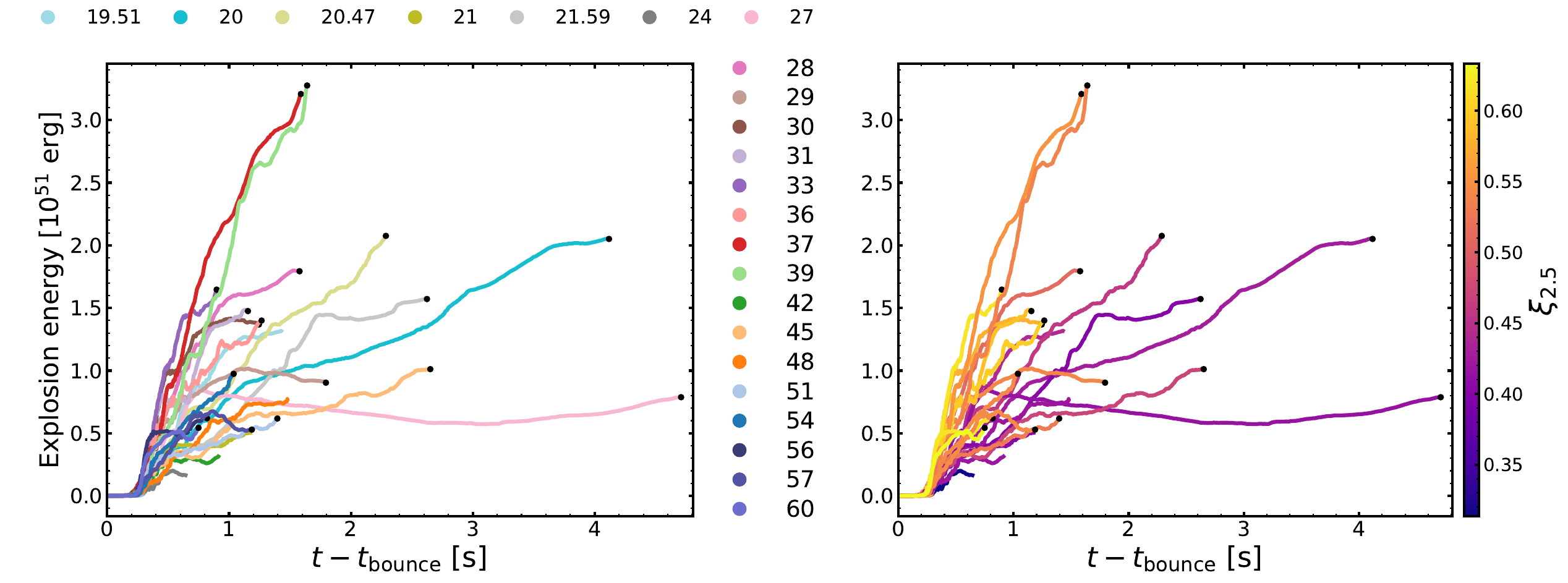}
  \end{adjustwidth}
  \caption{Diagnostic explosion energy, $E_\mathrm{diag}$, explicitly labeled (left), and color-coded by compactness (right). Large variations are seen in how the explosions develop, reflecting the wide range of progenitor structures, supplemented by the stochastic dynamical nature. Immediately after shock revival, the explosion energy tends to build faster with higher compactness. On the other hand, lower-compactness models can build their energy for a longer duration before BH formation, erasing any clear correlation between $E_\mathrm{diag}$ and  $\xi_{2.5}$ at BH formation. Moreover, details above and below the 2.5\,$M_\odot$ coordinate in the mass--radius relation Figure~\ref{fig:mass-radius} are of importance for the development of $E_\mathrm{diag}$, see text.}
  \label{fig:E_expl}
\end{figure*}

\begin{figure*}[ht!]
    \centering
    \includegraphics[width=1.5\columnwidth]{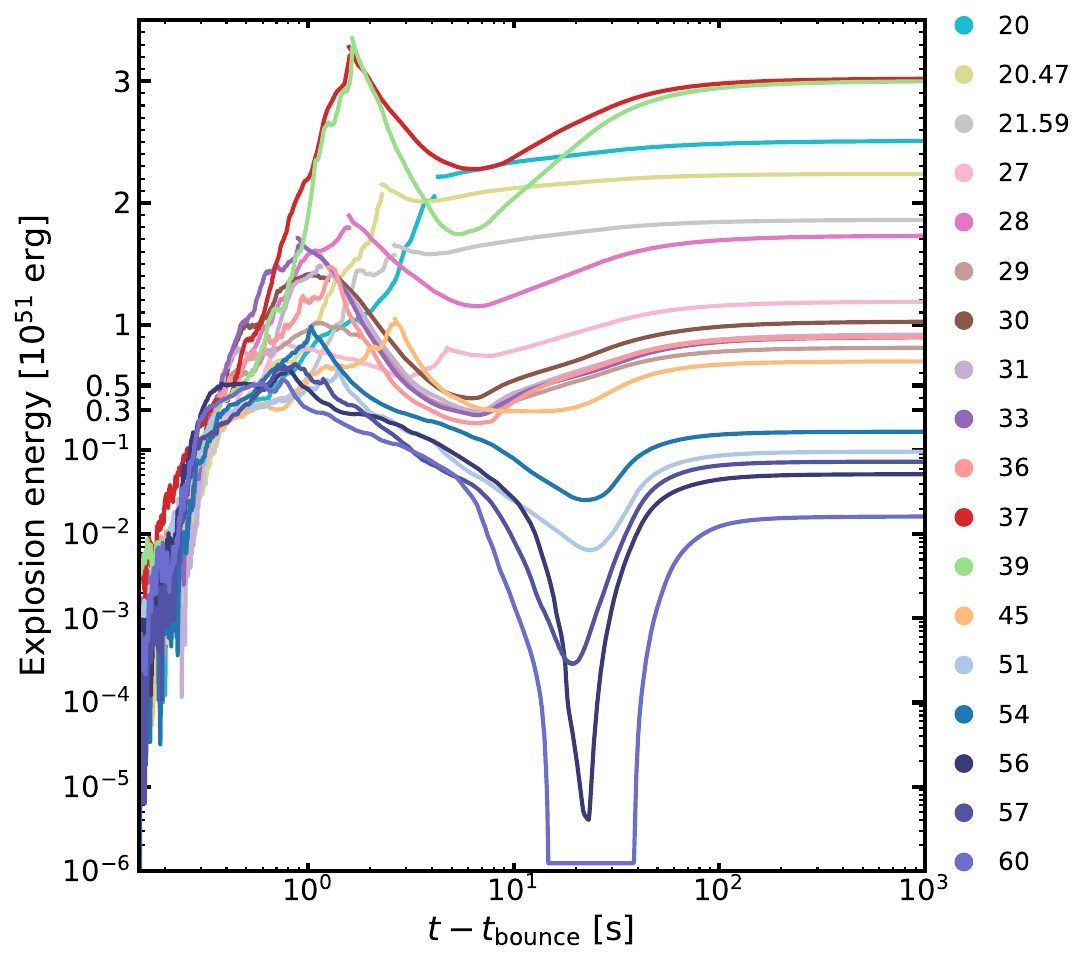}
    \caption{Diagnostic explosion energy curves (Equation~\ref{eq:diag_expl}) for the models that form BHs, plotted on a logarithmic scale up to $0.3\times10^{51}$\,erg and on a linear scale above this value. The energy peaks around BH formation when the neutrino engine ceases, then decreases as the energy in the neutrino-heated bubbles is drained while they propagate through the overburden. It rises again after the shock has exited the helium-free CO core , reflecting the kinetic energy of bound material moving outward and its redistribution in a low density, low binding energy environment. We highlight three interesting regimes. (1) The high-mass progenitors in this model set ($\geq 51\,M_\odot$) have massive, compact CO cores, which translate into a large overburden and therefore a decline of several orders of magnitude in $E_\mathrm{diag}$ after BH formation. This decline reflects the almost complete drainage of excess thermal energy in their neutrino-heated bubbles. Nevertheless, the shocks survive, unbinding only the loosely bound outer hydrogen envelope as $E_\mathrm{diag}$ rises again. (2) The low mass progenitors ($\leq 27\,M_\odot$), exhibiting the smallest CO cores and long durations before BH formation. Therefore, the explosions face a small overburden and are close to the outer edge of their CO cores at BH formation, resulting in small losses and a nearly constant $E_\mathrm{diag}$ after BH formation. (3) The intermediate regime (28--45\,$M_\odot$), where parts of the neutrino-heated regions are drained, but not completely, resulting in a moderate decrease and subsequent increase in $E_\mathrm{diag}$ after BH formation.} 
    \label{fig:E_explPBH}
\end{figure*}

\subsection{Explosion dynamics}
\subsubsection{Before Black Hole Formation}

\label{sect:explosion_dynamics}
Figure~\ref{fig:E_expl} shows the evolution of the diagnostic explosion energy ($E_\mathrm{diag}$). In the left panel, the simulations are explicitly labeled, while in the right panel they are color-coded by their compactness. We define the diagnostic explosion energy as \citep{buras:06b,mueller:17},
\begin{equation} \label{eq:diag_expl}
    E_\mathrm{diag} = \int_{e_\mathrm{tot},\, v_\mathrm{r} > 0} \rho e_\mathrm{tot} \, dV,
\end{equation}
with the total specific energy given by,
\begin{equation} \label{eq:etot}
    e_\mathrm{tot} = \epsilon_\mathrm{int} + \frac{v^2}{2} + \Phi,
\end{equation}
where $v_\mathrm{r}$ is the radial velocity, $\rho$ the density, $dV$ a volume element, $\epsilon_\mathrm{int}$ the internal energy, $v$ the fluid velocity, and $\Phi$ the gravitational potential. Here, $\epsilon_\mathrm{int}$ denotes the specific internal energy measured relative to the zero-temperature reference state at the same density and electron fraction,
\begin{equation}
    \epsilon_\mathrm{int} = \epsilon(\rho,T,Y_e) - \epsilon(\rho,0,Y_e).
\end{equation}
For the nuclear EOS, this removes the EOS-dependent rest-mass offset from the energy entering $e_\mathrm{tot}$. After BH formation, when the Helmholtz EOS is used throughout the domain, we do not apply an analogous subtraction and instead approximate the zero-temperature specific internal energy as negligible in the calculation.

Reflecting the wide progenitor range sampled and the stochasticity of the neutrino engine, the explosions exhibit large variations in their development (Figure~\ref{fig:E_expl}). Higher-compactness models tend to increase their explosion energy more rapidly, at least during the first few hundred milliseconds after shock revival, as expected since accretion/infall fuels the neutrino-driven engine \citep{nakamura:15, burrows:24, boccioli:25}. Despite a lower $E_\mathrm{diag}$ slope, lower compactness models can end up with a larger explosion energy at BH formation since the neutrino engine is active for a longer duration. Therefore, there is no clear trend between the explosion energy at BH formation and the compactness parameter. $E_\mathrm{diag}$ reaches $0.5$--$3.3 \times 10^{51}$\,erg in the BH forming models.

Due to the connection between mass-infall rate and explosion energy \citep{nakamura:15, bollig:21, burrows:24, boccioli:25}, details of the progenitor structure matter for the explosion development. Moreover, when shock revival sets in only after material from relatively large mass coordinates has accreted, less time remains for continued accretion before BH formation. Some of the highest-mass progenitor models initially show a rapid rise in explosion energy during the first few hundred milliseconds, but then their $E_\mathrm{diag}$ curves roll over. These progenitors (e.g. 60\,$M_\odot$, 57\,$M_\odot$, 54\,$M_\odot$, 51\,$M_\odot$) are very compact and have high heating rates at shock revival, but revival occurs only after accretion has reached relatively large mass coordinates, where the mass--radius curves bend and become shallower (see Figure~\ref{fig:mass-radius}). The resulting decrease in accretion rate is reflected both in the neutrino heating and in the rollover of $E_\mathrm{diag}$. Stars around the second compactness peak (e.g. 36\,$M_\odot$, 33\,$M_\odot$, 31\,$M_\odot$, 30\,$M_\odot$) for which shock revival occurs at slightly smaller mass coordinates and whose mass--radius profiles do not exhibit such sharp bends, also show a rapid rise in $E_\mathrm{diag}$ and do not roll over as early. The 37\,$M_\odot$ and 39\,$M_\odot$ models tower over the others in terms of diagnostic explosion energy, with $\sim4 \times 10^{51}$\,erg at BH formation. Their shocks expand earliest (both in time and mass-coordinate) due the early accretion of a sharp compositional interface (see Figure~\ref{fig:shock_radius}). With an early explosion and similar mass--radius slopes between 2-3\,$M_\odot$ (Figure~\ref{fig:mass-radius}) as the aforementioned models around the second compactness peak, they accrete a large amount of material at a high rate between shock revival and BH-formation, efficiently converting the binding energy of the accreting mass shells into explosion energy. Models such as the 21.59\,$M_\odot$ and 20\,$M_\odot$ exhibit relatively shallow mass–radius profiles (Figure~\ref{fig:mass-radius}), resulting in a modest yet prolonged accretion phase prior to BH formation and a slow, sustained buildup of explosion energy. Another interesting feature is seen in some models, where $E_\mathrm{diag}$ steadily decreases after a few hundred milliseconds. This is particularly evident in the 27\,$M_\odot$ model after $\sim750$\,ms. It highlights the dynamics of the expanding ejecta (with positive $e_\mathrm{tot}$ in Eq.~\ref{eq:etot}), which transfer energy to the overlying and surrounding bound layers when the infall rate and energy injection are low.
\subsubsection{After Black Hole Formation}
Before presenting how the explosions evolve after BH formation, we briefly recall the dynamics leading up to it. For these relatively high-compactness progenitors, the simultaneous development of explosion and accretion regulates the explosion morphology into a structure with a strong dipolar component. Perhaps this is also an emerging trend even in 3D simulations \citep{chan:18, burrows:23, burrows:25}; the initially stalled, yet dynamic shock front stochastically expands in one direction, seeding a sustained accretion funnel in the perpendicular direction. A collar-like accretion structure emerges, with outflows directed toward its openings. In 2D, however, the imposed axisymmetry can exaggerate this dipolar configuration and bias its orientation toward the symmetry axis. Despite large variations across the models, a common trend emerges in the simulations presented here. As the explosion develops, typically along the polar directions and often more strongly on one side, the infalling material is funneled toward the equatorial plane. During the first few seconds, the dominant contribution to $E_\mathrm{diag}$ arises from the neutrino-heated outflows. These bubbles contain high internal energy and are therefore more prone to being temporarily unbound than the outward moving material outside them, such as the material behind the shock in the equatorial plane.

For the models that form BHs, Figure~\ref{fig:E_explPBH} includes the explosion energy curves after BH formation, displayed on a logarithmic scale up to $0.3\times10^{51}$\,erg and on a linear scale above this value. We see the characteristic shape of $E_\mathrm{diag}$ \citep{chan:18, chan:20, rahman:22, eggenberger_andersen:25, sykes:25}, which, after the peak around BH formation, decreases due to energy transfer from the unbound neutrino heated bubbles behind the shock-front to surrounding and overlying regions, without being replenished by the neutrino engine from below. The decrease is primarily expected because of the binding energy associated with the overburden. $E_\mathrm{diag}$ rises again after $\sim5$--30\,s and ultimately settles on a final value for the explosion energy. Whether $E_\mathrm{diag}$ is the best measure for tracking the explosion before its properties are saturated is up for debate. Nevertheless, the evolution of $E_\mathrm{diag}$ offers insightful  information on the dynamics leading up to the stage when the energy redistribution has largely completed. In \citet{eggenberger_andersen:25}, we found the rise in $E_\mathrm{diag}$ to be associated with the shock exiting the CO core. Both the density and the binding energy of the material outside the helium-depleted CO core decline rapidly. Once the shock is in a region where they are low enough, the expanding material begins unbinding the material it encounters ahead. In this process, it is the kinetic energy of bound, outward-moving material that is responsible for the rise in $E_\mathrm{diag}$. We highlight that for BHSNe in which sizable regions of the neutrino heated bubbles remain unbound toward the edge of the CO core (models in the 20--45\,$M_\odot$ range), the rise in $E_\mathrm{diag}$ can generally be attributed to the bound, outward-moving material behind the shock in the equatorial region. 

The fall and subsequent rise of $E_\mathrm{diag}$ after BH formation is especially pronounced in the most massive progenitor models (the models above 51\,$M_\odot$), and exhibit similar behavior to models studied by \citet{sykes:25}. Their 1--5 order of magnitude decreases in $E_\mathrm{diag}$ indicate that nearly all excess thermal energy is drained from their neutrino-heated bubbles as they propagate toward the outer boundary of the CO core. In these models, nearly no material is unbound at the time when $E_\mathrm{diag}$ reaches its minimum. Nevertheless, their shocks survive, and the kinetic energy associated with the momentum of the outward-moving material behind the shock is redistributed, resulting in the rise of $E_\mathrm{diag}$ after $\sim20$\,s and final explosion energies between $\sim 10^{49}$--$10^{50}$\,erg.  As we will show in Section~\ref{sec:remnant_properties}, only the most loosely bound outer envelope is ejected in these models, whereas more of the stellar interior is ejected in the more energetic explosions.  

The low-mass models, 27\,$M_\odot$, 21.59\,$M_\odot$, 20.47\,$M_\odot$, and 20\,$M_\odot$, have remarkably weak declines in $E_\mathrm{diag}$ after BH formation. Instead, $E_\mathrm{diag}$ remains roughly constant in the critical seconds after BH formation. We show in Section~\ref{sec:remnant_properties} that the shocks in these models are closer to their outer CO core boundary at BH formation, and the energy in their neutrino-heated bubbles outweighs the overburden at BH formation. Thus, less energy redistribution occurs between bound and unbound material after BH formation, and the majority of the energy in the neutrino-heated bubbles survives the passage toward shock-breakout. 

Finally, we highlight that the final explosion energies are roughly ordered by the diagnostic explosion energy at BH formation and that the diverse set of final explosion energies will likely translate into a diverse set of transients from BHSNe.

\begin{figure}[ht!]
    \centering
    \includegraphics[width=\columnwidth]{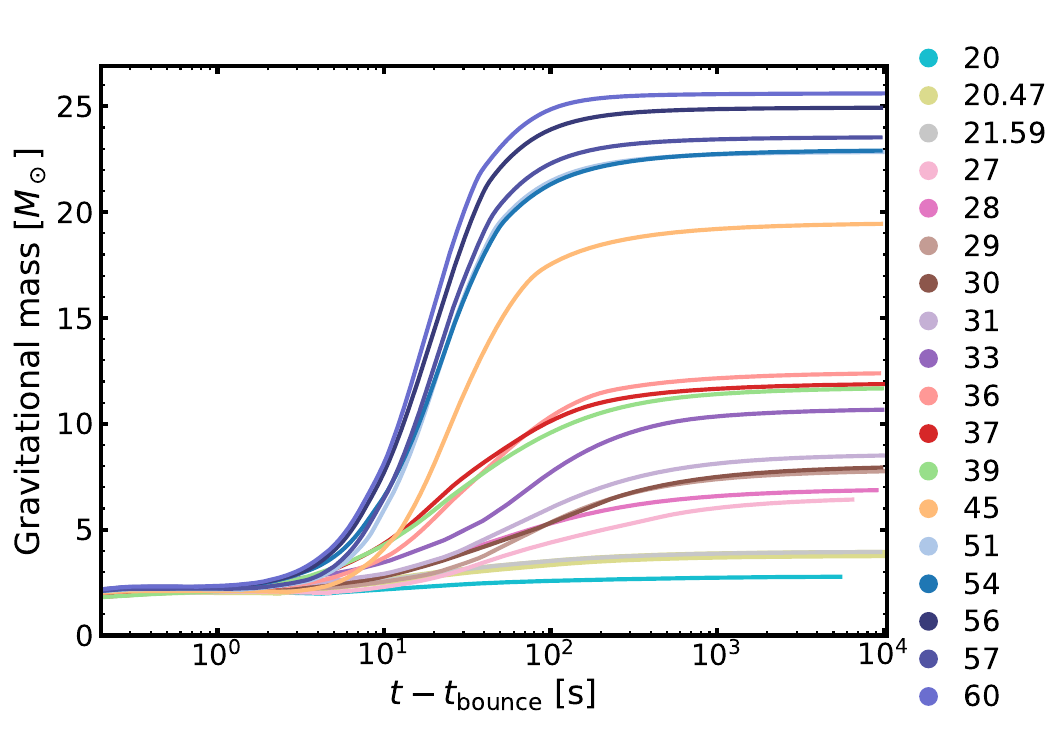}
   \caption{
Gravitational mass of the compact object for the BH-forming models. More massive progenitors leave more massive BHs behind because they possess more massive and more compact CO cores. Consequently, the overburden ahead of the shock at BH formation increases with CO-core mass (see Figure~\ref{fig:binding_energy}), and more mass accretes onto
the BH remnant.
}\label{fig:M_grav}
\end{figure}

\begin{figure}[ht!]
    \centering
    \includegraphics[width=\columnwidth]{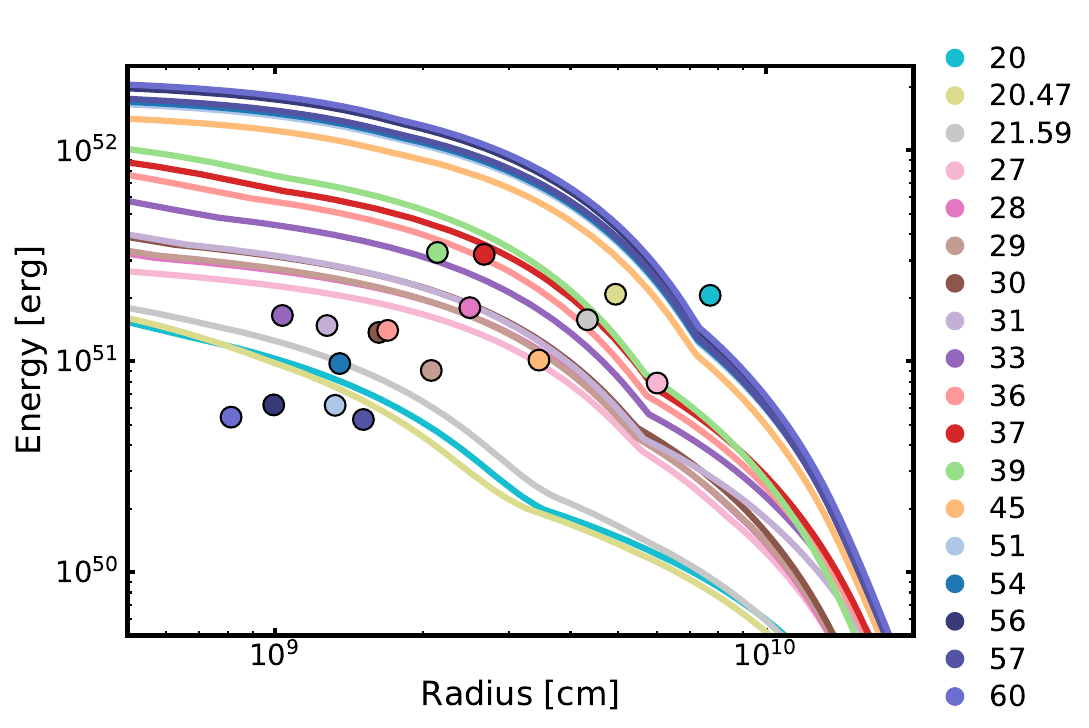}
    \caption{The overburden, i.e., the energy required to bring the total energy of the overlying material to zero, as a function of radius. The dots are $E_\mathrm{diag}$ at BH formation, placed at the radius of the shock at BH formation. Generally, the more massive the progenitor and its CO-core mass, the larger the disparity between $E_\mathrm{diag}$ and the overburden at BH formation. Four models have an $E_\mathrm{diag}$ higher than their overburden at BH formation: 27\,$M_\odot$, 21.59\,$M_\odot$, 20.47\,$M_\odot$, and 20\,$M_\odot$. These are the models with roughly constant $E_\mathrm{diag}$ after BH formation; the neutrino-heated bubbles easily survive the passage towards shock-breakout in these models.  } 
    \label{fig:binding_energy}
\end{figure}

\begin{figure*}[ht!]
    \centering
    \includegraphics[width=\linewidth]{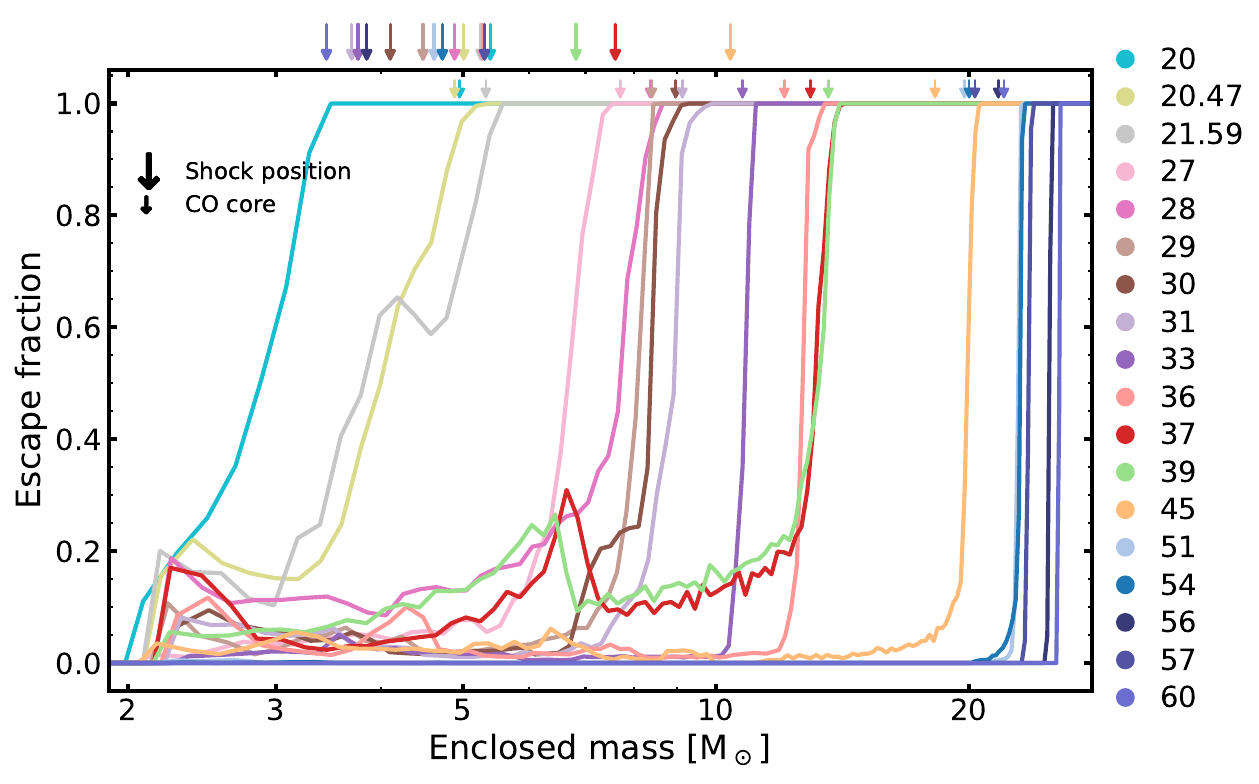}
    \caption{The fraction of each spherical mass shell at BH formation that ultimately escapes the BH and contributes to the ejecta. The long arrows denote the position of the shock in mass coordinate at BH formation, while the short arrows indicate the outer boundary of the helium-depleted CO core. For models of 51\,$M_\odot$ and above, only loosely bound hydrogen envelope is ejected, marked by the sharp transition from $\sim0$ to 1. For all other models, it is not possible to apply a simple spherical mass cut to predict the partitioning of the star into BH mass and ejecta mass. Instead, a fraction of each mass shell outside the BH contributes to the ejecta. In the intermediate mass regime (28--45\,$M_\odot$), however, the escape fraction approaches unity near the helium-depleted CO core boundary. In the lower-mass models (20\,$M_\odot$, 20.47\,$M_\odot$, and 21.59\,$M_\odot$), the shocks have already propagated close to the outer edge of the CO core due to the combination of late BH formation and small CO cores. As a result, most material immediately behind the shock and nearly all material ahead of it at BH formation is ejected.  } 
    \label{fig:probability_of_escape}
\end{figure*}

\begin{figure*}[ht!]
    \centering
    \includegraphics[width=\linewidth]{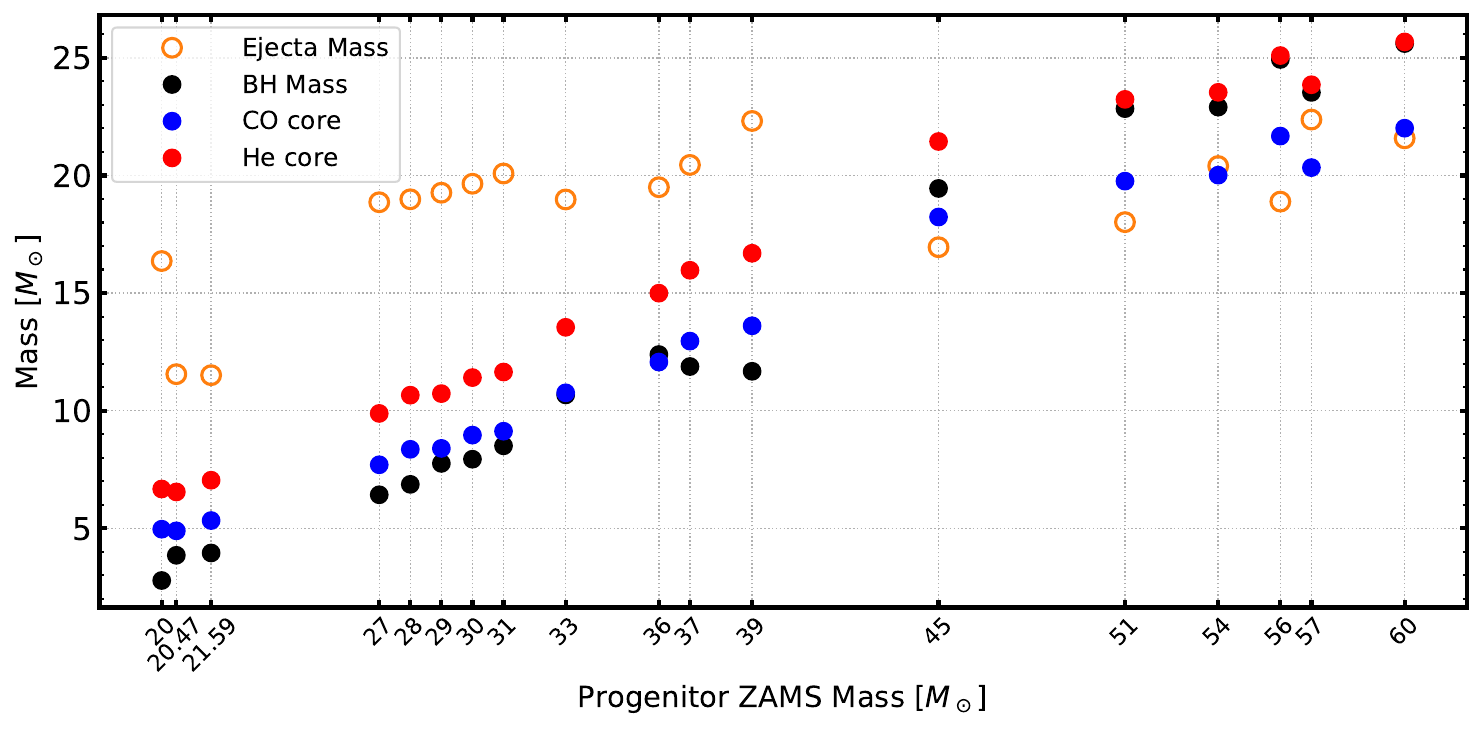}
    \caption{The final BH mass (black dots), the ejecta mass (orange circles), the helium-depleted CO-core mass (blue dots), and the hydrogen-depleted He-core mass (red dots). The remnant mass increases approximately linearly with ZAMS progenitor mass, owing to the approximately linear relation between the CO-core mass and ZAMS mass in this progenitor suite; higher mass models have more mass that is simultaneously more tightly bound. For the highest-mass models, 51\,$M_\odot$ and above, the remnant mass is close to that of the hydrogen-depleted He-core mass because of the high overburden, resulting in a low-energy explosion that only unbinds the envelope. Models below 39\,$M_\odot$ eject more mass than just the envelope since the neutrino-heated bubbles survive the binding energy of the core and bring material from depths to the stellar surface in the explosion. In particular, the low-mass models (20\,$M_\odot$, 20.47\,$M_\odot$, and 21.59\,$M_\odot$) exhibit a significant deviation between remnant mass and CO-core mass, with remnant masses amounting to only 56--79\,$\%$ of the helium-depleted CO-core mass.} 
    \label{fig:BH_ejecta_mass}
\end{figure*}

\subsection{BH remnant properties}\label{sec:remnant_properties}
After BH formation, the absence of a central pressure source allows material to freely accrete into the BH. Furthermore, as the shocks move out, not all of the overlying layers are stripped. Figure~\ref{fig:M_grav} shows the gravitational mass growth of the BHs, which accrete the majority of their mass in the first 100s of seconds, and have approximately reached their final mass after $\sim 1000$\,s. Final BH masses range from $\sim3$\,$M_\odot$ to $\sim26$\,$M_\odot$. The mass growth is generally ordered by progenitor mass, with higher-mass progenitors leaving more massive BHs behind. We attribute this to their more massive and more compact CO cores, which translate into a larger overburden ahead of the shock at BH formation. In this progenitor set (Section~\ref{sect:progenitor}), reduced mass loss leads to a CO-core mass that increases roughly monotonically with ZAMS mass, a behavior that does not generally hold across all stellar evolution models. 

The overburden is shown in Figure~\ref{fig:binding_energy}, where each line represents the energy required to bring the total energy of the overlying material to zero, as a function of radius. The dots are $E_\mathrm{diag}$ at BH formation, placed where the mean shock radius is located at that time. Only four models have an $E_\mathrm{diag}$ higher than their overburden at BH formation: 27\,$M_\odot$, 21.59\,$M_\odot$, 20.47\,$M_\odot$, and 20\,$M_\odot$. These are the models without any significant decline in $E_\mathrm{diag}$ immediately after BH formation; the majority of the energy in the neutrino heated bubbles survives for these models, and the final explosion energy is close to the energy when the neutrino engine ceased, leaving behind remarkably low-mass remnants. The remaining models have a lower $E_\mathrm{diag}$ than the overburden at BH formation. In \citet{eggenberger_andersen:25}, we showed that the additional kinetic energy of bound, outward-moving material did not raise the total energy budget above the overburden, prompting the question of why explosions still occur despite the available energy being lower than the overburden at BH formation. We reasoned as follows: the majority of the explosive energy is concentrated in neutrino-heated bubbles rather than being distributed isotropically over the sphere. This anisotropic morphology allows much of the overburden to be circumvented, as the bubbles propagate outward without encountering the full overburden at all solid angles. This picture is consistent across the models presented in this paper. For the most massive progenitors (51\,$M_\odot$ and above), however, the overburden is sufficiently large that any significant energy stored in the bubbles is lost, leaving only a shock that propagates outward in mass coordinate and eventually only unbinds envelope material.

Figure~\ref{fig:probability_of_escape} shows, for each spherical mass shell at BH formation, the fraction of mass that ultimately escapes the BH and contributes to the ejecta. The long arrows indicate the shock position at BH formation, while the short arrows show the mass coordinate of the outer edge of the helium-depleted CO core. The shocks and heated bubbles in the high-mass progenitor models (51\,$M_\odot$ and above) have many solar masses of material to pass through before reaching the envelope. Since the explosion energy is drained in the process, only material outside of the CO core is ejected. Specifically for these models, we note that the boundary between accreted and ejected material is quite sharp and spherical, which results in a sharp jump from 0 to 1 in the escape fraction at a certain mass coordinate. Interestingly, this mass coordinate corresponds to the outer edge of the hydrogen-depleted He core. In other words, only the loosely bound hydrogen envelope is ejected. In contrast, for models with a ZAMS mass below 51\,$M_\odot$, it is not possible to simply apply a mass cut to predict the partitioning of the star into the BH mass and the ejecta mass, and the aforementioned boundary is highly aspherical. Generally, some fraction of each mass shell outside the BH contributes to the ejecta. In the intermediate mass regime (28--45\,$M_\odot$), however, the escape fraction reaches one in close proximity to the helium-depleted CO core boundary. The shocks in the lower-mass models, the 20\,$M_\odot$, 20.47\,$M_\odot$, and 21.59\,$M_\odot$, have already reached far towards the edge of the CO core due to the combination of late BH formation and small CO cores. As a result, most material immediately behind and nearly all in front of the shock at BH formation is ejected. 

Figure~\ref{fig:BH_ejecta_mass} sums up this section by showing the remnant mass (black dots), the ejecta mass (orange circles), the helium-depleted CO-core mass (blue dots), and the hydrogen-depleted He-core mass (red dots) for each model. A key feature of this diagram is how the BH masses increase approximately linearly with ZAMS progenitor mass. This feature stems from the approximately linear relation between the CO-core mass and ZAMS mass in this progenitor suite; higher mass models have more mass that is simultaneously more tightly bound. For the highest-mass models, 51\,$M_\odot$ and above, most of the energy in the neutrino-heated bubbles is drained, leaving only a surviving shock that unbinds the hydrogen envelope, such that remnant mass is very close to the hydrogen-depleted He-core mass, consistent with \citet{sykes:25}. The close agreement between the remnant mass and the He-core mass breaks down for models below 51\,$M_\odot$. In these models, parts of the neutrino-heated bubbles survive the binding energy of the core, ejecting more mass than just the envelope. In the model range of (28--45\,$M_\odot$), a closer predictor of the remnant mass is the helium-depleted CO-core mass. This agreement is nearly exact for the 33\,$M_\odot$, which is likely reflecting the early BH formation time \citep{eggenberger_andersen:25} due to its location at the top of a compactness peak (Figure~\ref{fig:compactenss_sample}). The remnant masses are, however, generally smaller than the helium-depleted CO-core masses. The difference between the remnant mass and the CO-core mass is particularly wide for the two most energetic models, 37\,$M_\odot$ and 39\,$M_\odot$. The remnant masses for the lowest-mass models, 20\,$M_\odot$, 20.47\,$M_\odot$, and 21.59\,$M_\odot$, also exhibit a significant deviation from their CO-core masses, with remnant masses between 56--79\,$\%$ of the helium-depleted CO-core mass. BHs of $\sim3.9$\,$M_\odot$ are produced in two of these models (20.47\,$M_\odot$, and 21.59\,$M_\odot$), while the remnant mass is $\sim2.8$\,$M_\odot$ for the 20\,$M_\odot$ model. All three BHs fall within the so-called lower mass-gap \citep[e.g.][]{corral-santana:16, xing:25}.  Our method of calculating the gravitational mass via Equation\,\ref{eq:mgrav} is approximate and likely underestimates the gravitational mass, particularly for PNSs that collapse to a BH after several seconds. The underestimation can be viewed as a fraction of the mass we subtract from the baryonic mass at BH formation due to the neutrino emission until that time, and if we assign a very cautious fraction of 50\,\%, the resulting errorbar on the final gravitational mass is $\sim6$\,\%. This does not alter the picture of BHSNe as a candidate mechanism for populating the mass-gap. We remark that we have not simulated into the regime where the reverse shock may bind some material that ultimately falls back. In \citet{eggenberger_andersen:25}, we saw no significant mass accretion owing to the reverse shock, but this may not generalize to every progenitor.

Above 27\,$M_\odot$, the ejecta mass lies within a band of $\sim17$--$22$\,$M_\odot$. Higher mass models have more envelope mass but less ejecta from deep layers, while the opposite is true for lower mass models. This balancing effect results in a roughly constant ejecta mass for these progenitor models in this ZAMS mass range. How the envelope mass and the CO-core mass vary with ZAMS mass is highly dependent on the stellar evolution model and the mass-loss rate employed. This can in part be seen by comparing the lower ejecta mass in the 21.59\,$M_\odot$ and 20.47\,$M_\odot$ models to that of the 20\,$M_\odot$ model. The latter has a lower mass-loss rate than the former two (see details in Section~\ref{sect:progenitor}), and therefore has a more massive envelope to eject. We expect more pronounced differences for higher ZAMS mass models at different mass-loss rates.

\section{Discussion}\label{sec:discussion}

\subsection{The Black Hole Mass Distribution}
\begin{figure}[ht!]
    \centering
    \includegraphics[width=\columnwidth]{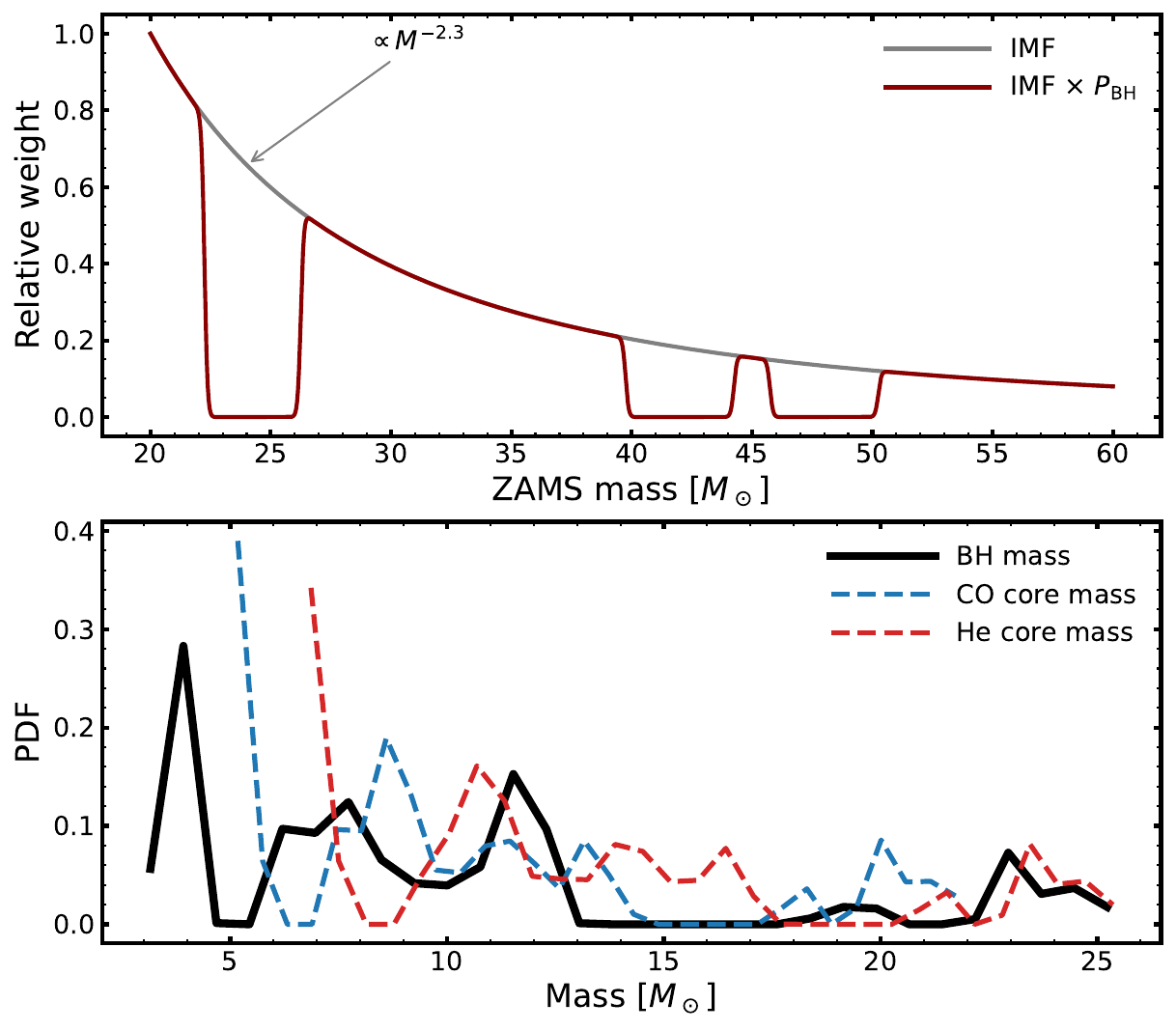}
    \caption{BH mass distribution (bottom panel, black line) obtained by convolving the progenitor mass--remnant mass relation with the function  \(\phi(M)\,P_{\rm BH}(M)\), where \(\phi(M)\propto M^{-2.3}\) is the initial mass function (top panel, gray line), and where \(P_{\rm BH}(M)\) is representing regions in ZAMS mass where we do not explicitly obtain BHs in the simulations (top panel, red line). The dashed lines in the bottom panel is the distributions obtained by convolving the helium-depleted CO-core mass and the hydrogen-depleted He-core mass with  \(\phi(M)\,P_{\rm BH}(M)\). At low masses, $\sim$3--8\,$M_\odot$, the BH distribution is downshifted from the CO-core distribution. The pronounced peak around 12\,$M_\odot$ corresponds to the highly energetic 37--39\,$M_\odot$ region with a large amount of ejected mass. In the high-mass end, the bump around $\sim23$\,$M_\odot$ is very well approximated by the He-core distribution.} 
    \label{fig:BH_mass_dist}
\end{figure}
Given accurate physics input into a simulation, it is the structure of the inner $\sim3$\,$M_\odot$ of the progenitor that dictates the explosion initiation, the subsequent accretion rate and BH formation time -- the key ingredients of BHSNe. With the uncertainties in the relationship between a newly formed star and its final structure, such that different stellar evolution models yield different ZAMS mass--compactness relations, it is premature to draw firm conclusions regarding a particular ZAMS mass and its likelihood of ending in a BHSN. Nevertheless, since several works in the literature have collectively shown that BHSNe are viable theoretical outcomes of core-collapse events, it is beneficial to begin the exercise of estimating their potential astrophysical impact and to allow future work to update the picture. To this end, we show the resulting BH mass distribution from the sample of BHSNe obtained in this study in Figure~\ref{fig:BH_mass_dist} (bottom panel). It is calculated by linearly interpolating the ZAMS mass--BH remnant relation (Figure~\ref{fig:BH_ejecta_mass}), and convolving this information with an initial mass function. For the initial mass function, we use an exponent of 2.3 \citep{kroupa:21}, and suppress regions in ZAMS mass where BHs are not formed in this study within the simulation time (top panel). Common assumptions for estimating the final BH mass in failed SNe are that it equals either the total stellar mass at core collapse, the He-core mass, or the CO-core mass. To assess whether the latter two approximations are also applicable to BHSNe, we include the corresponding distributions obtained by assuming that the remnant mass equals the hydrogen-depleted He-core mass or the helium-depleted CO-core mass (bottom panel, dashed lines). A comparison of the distributions in the bottom panel shows that the main features of the BH mass distribution can still be interpreted from the shapes of the CO-core and He-core distributions, although the resulting BH mass distribution corresponds to neither one. At low masses, around 3--8\,$M_\odot$, the BH distribution is shifted downward relative to the CO-core distribution since the lower-mass progenitors make BHs that are up to $\sim44\,\%$ less massive than their CO-core mass. The pronounced peak near 12\,$M_\odot$ originates from the highly energetic 37--39\,$M_\odot$ progenitors, which eject a large amount of mass, effectively flattening out the progenitor--remnant relation. At the high-mass end, the bump around 23\,$M_\odot$ is reproduced well by the He-core distribution. The simulations in this study suggest that, for CO-cores more massive than about $\sim20$\,$M_\odot$, the remnant mass is reasonably approximated by the hydrogen-depleted He-core mass in a BHSN context.

\subsection{The Impact of Resolution on Late-time PNS Modelling}\label{sect:resolution}
In this section, we highlight a challenge in late-time numerical modeling of PNSs. As a PNS accumulates mass and cools, the density gradient at its surface steepens. At some stage after a second of evolution post bounce, with a grid spacing of 381\,m, this gradient can become so steep such that only $\sim3$ zones cover the order of magnitude in density between $10^{11}$\,g\,cm$^{-3}$--$10^{12}$\,g\,cm$^{-3}$. Once the PNS surface becomes under-resolved, we find that the configuration of the PNS is not spatially converged, resulting in a central density that gradually stagnates and may behave unexpectedly if the simulation continues. This may then affect the BH formation time. If a higher resolution is employed in this phase, the PNS surface immediately contracts and the central density gradually shifts toward a more accurate value as the PNS relaxes into slightly more compact configuration. This can be seen in the central density figure (Figure~\ref{fig:central_density}) for the $20\,M_\odot$ simulation around $\sim3.6$\,s, when the resolution was changed from $\sim381$\,m to $\sim191$\,m. The central density evolution in this model is comparable to that of the $27\,M_\odot$ model, which is run with $\sim191$\,m resolution all the way from $\sim0.6$\,s after bounce, and was tested for spatial convergence after 4\,s using $\sim95$\,m. Since the PNS evolution of the $20\,M_\odot$ model is similar to that of the $27\,M_\odot$ model, and since the evolution and BH formation on the mass--entropy diagram is reasonable  (Figure~\ref{fig:mass_entropy}), we expect that rerunning the $20\,M_\odot$ model at a higher initial resolution would yield a similar BH formation time. We therefore retain this simulation in our presented set of BHSN models. The $21.59\,M_\odot$ model also entered into a regime where the PNS surface was unresolved. For this model, we restarted the simulation with a higher resolution at $\sim1.2$\,s, a time early enough such that the numerical solution was unaffected by the resolution. 

Since long-time multidimensional simulations of PNSs are becoming more prevalent, careful attention to these effects can help circumvent the need for late-time restarts and improve the reliability of BH formation predictions.

\subsection{2D vs 3D: A New Instance of a BHSN in 3D}

Core-collapse events and SNe are fundamentally three-dimensional phenomena. The results of this study strongly indicate that the CO-core mass is the dominant factor setting the relationship between progenitor mass and remnant mass. At the same time, individual models, such as the 37\,$M_\odot$ and 39\,$M_\odot$ simulations, demonstrate that the details of the explosion development also matter, and such details are more accurately captured by three-dimensional models. While BH formation after shock revival has been reported in multiple three-dimensional simulations \citep[e.g.,][]{kuroda:18, walk:20, pan:21, powell:21}, the evolution is not always followed beyond BH formation, which can make the final outcome difficult to determine. To our knowledge, the existing three-dimensional simulations of BHSNe are those of \citet{chan:18, chan:20}, who modeled a 40\,$M_\odot$ zero-metallicity progenitor and found that the entire He core was eventually accreted, and those of \citet{burrows:23, burrows:25}, who found that a different 40\,$M_\odot$ progenitor with $\xi_{2.5}=0.54$ left behind a BH of \(\sim9\,M_\odot\), while a 19.56\,$M_\odot$ progenitor with $\xi_{2.5}=0.45$ produced a BH remnant of 3.12\,$M_\odot$. A remnant mass of 3.12\,$M_\odot$ for the  19.56\,$M_\odot$ model in Burrows et al. is in good agreement with the remnants of the low-mass progenitors in this 2D study (see Figure~\ref{fig:BH_ejecta_mass}).\\ 

We extend this discussion by presenting a previously unpublished 3D BHSN simulation \citep[][in preparation]{liubov:26}, evolved with the FLASH code without initial rotation. The progenitor is a stripped 39\,$M_\odot$ progenitor from \citep{aguilera-dena:20}, which is more compact than any of the 2D models presented in this paper, with $\xi_{2.5}=0.75$. The EOS employed is SRO 0.95 \citep{schneider:19}, which generally delays the BH formation time compared to using the SFHo EOS. Since the CO-core mass is very large, of roughly $28\,M_\odot$, it behaviorally falls into the same category as the high-mass category presented in Section~\ref{sec:results}, where nearly the entire CO core accretes into the BH. Since it is a stripped progenitor, only $\sim1\,M_\odot$ of material is ejected with a final explosion energy of $\sim0.3\times10^{51}$\,erg. At BH formation, two pronounced neutrino heated bubbles are extended in opposite direction, not unlike the morphology in many of the axisymmetric models. We have also simulated this progenitor in 2D using an otherwise identical setup, and provide a direct comparison of the evolution leading up to BH formation.

The early evolution is very similar in 2D and 3D. Core bounce occurs at the same time, shock revival follows at nearly identical times, and the shock expansion during the early post-revival phase is comparable. The overall shock evolution also remains closely matched at later times. The primary difference lies in the PNS evolution. For roughly the first \(\sim150\) ms after shock revival, the PNS radii differ by only a few kilometers. Thereafter, the 2D PNS contracts more rapidly, whereas the 3D PNS contracts more gradually and remains stable for much longer. Nevertheless, both models collapse at nearly the same PNS radius, \(\sim25\) km in 2D and \(\sim24\) km in 3D. As a result, BH formation occurs much earlier in 2D, at \(t_{\rm BH}^{\rm 2D}\simeq620\)\,ms after bounce, than in 3D, where collapse is delayed until \(t_{\rm BH}^{\rm 3D}\simeq1\)\,s. Since the resolution is fairly coarse in the 3D simulation ($\sim600$\,m) we likely are unresolving the surface, as discussed in Section~\ref{sect:resolution}. While the resolution in the 2D simulation is the same, the added dimension (and in particular the lack of assumed axisymmetry) may lead to a different manifestation of resolution dependence.  We have performed two other 2D simulations at 1.5$\times$ and 2$\times$ resolutions and see a reduction of the black hole formation time by 20\% (down to $\sim 500$\,ms for the 2$\times$ case). Nevertheless, at \(t_{\rm BH}^{\rm 2D}\) the PNS mass in the 3D simulation is a few per cent less massive than in the 2D simulations, and the accretion rate has slowed down and even reversed, which may be an expected 3D feature \citep{mueller:15b, mueller:17}. With little to no accretion, a simulation on the mass--entropy diagram moves horizontally. Since the entropy decrease slows down with time, our analysis suggests that it can take several hundreds of milliseconds before the PNS has cooled enough to reach the critical threshold in PNS mass--entropy space (see Figure~\ref{fig:mass_entropy}).

A systematic study of the differences between 2D and 3D in the phase leading up to BH formation is warranted. Contributing toward this end, we have begun a 3D simulation of the same 30\,$M_\odot$ progenitor as the one presented in Section~\ref{sec:results}.

\section{Conclusions}\label{sec:conclusions}

We have presented a suite of 23 long-term axisymmetric core-collapse simulations of single-star progenitors from \citet{sukhbold:18}, spanning $19.51$--$60\,M_\odot$ and compactnesses of $0.31 \lesssim \xi_{2.5} \lesssim 0.63$. Using FLASH with energy-dependent M1 neutrino transport and the SFHo EOS, we followed the evolution from collapse through shock revival and, for BH-forming models, continued the post-BH evolution with an excision treatment until at least $5000$\,s, when the remnant and explosion properties had approximately asymptoted. Our main goal was to assess how broadly BHSNe arise across progenitor structure and how their explosions and remnant properties depend on that structure. Our main conclusions are as follows.

First, BHSNe emerge across a broad progenitor range in this model set. Of the 23 progenitors simulated, 18 form BHs after shock revival, with BH formation times between $\sim0.7$\,s and $\sim4.4$\,s after bounce. These BHSN outcomes occur across nearly the full ZAMS mass range considered and for compactnesses of approximately $0.40 \lesssim \xi_{2.5} \lesssim 0.63$. In this sense, our results extend previous BHSN studies, which have largely focused on very massive and highly compact progenitors, and show that BHSNe are not restricted to the most extreme stars.

Second, the compactness remains an important organizing parameter for BHSNe, but it is not sufficient on its own to determine the detailed evolution leading up to BH formation. Higher compactness generally leads to earlier BH formation and more rapid initial explosion-energy growth, reflecting higher accretion
rates and stronger neutrino heating. However, the detailed path to BH formation, the explosion energy
at BH formation, and the final remnant mass are all sensitive to the broader inner mass--radius structure
of the progenitor beyond the single compactness parameter $\xi_{2.5}$.

Third, the evolution of the PNS toward BH formation in these BHSN models is well interpreted in the baryonic mass--entropy plane. Although shock revival bends the evolutionary tracks relative to the nearly straight trajectories expected in failed SNe, BH formation still occurs close to the general relativistic maximum-mass curve $M^{\mathrm{max}}(s)$ for constant-entropy stars. This indicates that the collapse is still governed by the thermodynamic state of the PNS in a way that is broadly consistent with the physical picture established for failed SNe, despite the additional complexity introduced by multidimensional explosion dynamics and ongoing accretion.

Fourth, the explosion dynamics after BH formation show a common qualitative behavior but quantitatively diverse outcomes. Generally, $E_\mathrm{diag}$ peaks near BH formation, then declines as the energy stored in the neutrino-heated bubbles is redistributed to surrounding and overlying bound material after the neutrino engine is shut off. At later times, $E_\mathrm{diag}$ rises again once the shock reaches regions of sufficiently low density and binding energy, which occurs after the shock exits the helium-depleted CO core. The depth of the post-BH decline and the eventual final explosion energy depend strongly on the overburden and on how much of the neutrino-heated bubbles survive the propagation through the stellar interior.

Fifth, the final BHSN outcomes span a wide range of explosion energies and remnant masses. The final explosion energies range from $\sim2\times10^{49}$\,erg to $\sim3\times10^{51}$\,erg, implying that BHSNe may produce a diverse class of transients rather than a single characteristic observational signature. The BH gravitational masses span $\sim3$--$26\,M_\odot$ and generally increase with progenitor mass in this progenitor suite, reflecting the approximately monotonic increase of the CO-core mass with ZAMS mass in this model set. The lowest-mass BHSN progenitors in our sample produce BHs in or near the purported lower mass gap, while the highest-mass progenitors  leave behind massive BHs and only weak explosions that primarily unbind the hydrogen envelope.

Sixth, there is generally no single spherical mass coordinate that cleanly partitions the star into ejecta and remnant. For progenitors below $\sim51\,M_\odot$, the partition between accreted and ejected matter is highly aspherical, and some fraction of many mass shells outside the BH at formation contributes to the ejecta. Only in the highest-mass models does a relatively sharp, nearly spherical mass cut emerge, in which the explosion mainly ejects the loosely bound hydrogen envelope and the remnant mass lies close to the hydrogen-depleted He-core mass. In the intermediate-mass regime, the helium-depleted CO-core mass is a more useful, though still imperfect, predictor of the final BH mass.

Finally, our results highlight both the promise and the numerical challenges of long-term multidimensional PNS evolution. We find that insufficient resolution at late times can under-resolve the steep density gradient at the PNS surface, altering the central density evolution and potentially biasing the BH formation time. This issue becomes important after $\sim1$\,s post-bounce and deserves careful attention in future work. Since long-time multidimensional simulations of PNSs are becoming more prevalent, resolving these late-time structural gradients will be important for obtaining robust BH-formation predictions.

Taken together, our results support the picture that BHSNe are a viable and potentially common theoretical outcome among relatively compact massive stars. They occupy a distinct regime between ordinary successful SNe and failed explosions, with simultaneous explosion and accretion shaping both the remnant and the ejecta. If realized in nature, BHSNe may contribute to the stellar-mass BH distribution across a broad mass range, including the lower mass gap, while also producing a correspondingly broad phenomenology in explosion energies and observable transients. Because the explosion unbinds a portion of the stellar interior, the remnants of BHSNe are systematically less massive than those expected from failed SNe of similar progenitors.

Several important questions remain open. The present study is restricted to axisymmetry, and fully three-dimensional simulations will be needed to determine how robust the obtained remnant mass--CO-core mass relation is. Our comparison of two FLASH simulations, one in each dimensionality, indicate that the BH formation time may be delayed in 3D, but a systematic comparison of dimensional effects is warranted. Additionally, the observable signatures of BHSNe remain to be quantified. Future work should couple these explosion models to radiative-transfer and nucleosynthesis calculations in order to predict light curves, spectra, and ejecta yields, and thereby assess how BHSNe may be identified observationally and how strongly they contribute to chemical enrichment.

\begin{acknowledgments}
We thank Anders Jerkstrand, Jesper Sollerman, Eva Laplace, Jonah Miller, Kaustav Das, Philipp Podsiadlowski, Raphael Hirschi, and Matteo Bugli for valuable discussions that benefited this study. This work is supported by the Swedish Research Council (Project No. 2020-00452). The computations were enabled by resources provided by the National Academic Infrastructure for Supercomputing in Sweden (NAISS), partially funded by the Swedish Research Council through grant agreement no. 2022-06725. S.M.C. is supported by the U.S. Department of Energy, Office of Science, Office of Nuclear Physics, under award No. DE-SC0017955.
\end{acknowledgments}

\software{FLASH \citep{fryxell:00}, NuLib \citep{oconnor:15a}, Matplotlib \citep{hunter:07}, NumPy \citep{harris2020array}, SciPy \citep{2020SciPy-NMeth}, yt \citep{turk:11}, h5py \citep{collette_python_hdf5_2014}}

\end{document}